\documentclass[12pt]{article}
\usepackage{amsmath,cite,epsfig,a4,times,euler,euscript}
\usepackage{graphics}
\usepackage[usenames]{color}
\renewcommand{\baselinestretch}{1.1}
\newcommand{\myTitle}[1]{\begin{center}{\bf\Huge #1}\\[5ex]\end{center}}
\newcommand{\myAuthor}[1]{\begin{center}{\Large #1}\\[2ex]\end{center}}
\newcommand{\myAffiliation}[1]{\\[1ex]{\it\large #1}}
\newcommand{\myEmail}[1]{}
\newcommand{\myDate}{\begin{center}{\large\today}\\[5ex]\end{center}}
\newcommand{\myAbstract}[1]{\begin{center}\renewcommand{\baselinestretch}{1}{\bf Abstract}\\[2ex]\parbox{0.8\linewidth}{\small\hspace{15pt} #1}\end{center}\vspace{\baselineskip}}
\newcommand{\myReport}[1]{\hspace{\fill} #1}
\newcommand{\myPreprint}[1]{}
\newcommand{\myKeywords}[1]{}
\newcommand{\myFigure}[1]{\begin{figure}\begin{center}#1\end{center}\end{figure}}

\newcommand{\myScript}[1]{\EuScript{#1}}

\newcommand{\fudgeb}{\\[-0.7ex]}

\newcommand{\lid}[2]{#1\!\cdot\!#2}
\newcommand{\slashp}{p\hspace{-6.5pt}/}
\newcommand{\slashq}{q\hspace{-6.5pt}/}
\newcommand{\slashl}{\ell\hspace{-6.0pt}/}
\newcommand{\Appendix}[1]{Appendix~\ref{#1}}   
\newcommand{\Section}[1]{Section~\ref{#1}}

\newcommand{\Figure}[1]{Fig.~\ref{#1}}
\newcommand{\Equation}[1]{Eq.~(\ref{#1})}
\newcommand{\Ref}[1]{(\ref{#1})}
\newcommand{\ie}{{\it i.e.}}
\newcommand{\cf}{{\it cf.}}
\newcommand{\eg}{{\it e.g.}}
\newcommand{\Amp}{\myScript{A}}
\newcommand{\bram}[1]{\langle#1|}
\newcommand{\brap}[1]{[#1|}
\newcommand{\ketm}[1]{|#1]}
\newcommand{\ketp}[1]{|#1\rangle}
\newcommand{\bkmm}[2]{\langle#1|#2]}
\newcommand{\bkmp}[2]{\langle#1|#2\rangle}
\newcommand{\bkpm}[2]{[#1|#2]}
\newcommand{\bkpp}[2]{[#1|#2\rangle}
\newcommand{\srac}[2]{{\textstyle\frac{#1}{#2}}}
\newcommand{\Tr}{\mathrm{Tr}}
\newcommand{\imag}{\mathrm{i}}
\newcommand{\gqcd}{g_{\mathrm{s}}}
\newcommand{\xone}{x_1}
\newcommand{\xtwo}{x_2}
\newcommand{\GeV}{\mathrm{GeV}}

\begin{document}

\myReport{IFJPAN-IV-2012-12}
\myPreprint{}

\myTitle{%
Helicity amplitudes\\[0.5ex] for high-energy scattering%
}

\myAuthor{%
A.~van~Hameren, P.~Kotko and K.~Kutak%
\myAffiliation{%
The H.\ Niewodnicza\'nski Institute of Nuclear Physics\fudgeb
Polisch Academy of Sciences\\
Radzikowskiego 152, 31-342 Cracow, Poland%
\myEmail{hameren@ifj.edu.pl}
}
}

\myDate

\myAbstract{%
We present a prescription to calculate manifestly gauge invariant tree-level helicity amplitudes for arbitrary scattering processes with off-shell initial-state gluons within the kinematics of high-energy scattering.
We show that it is equivalent to Lipatov's effective action approach, and show its computational potential through numerical calculations for scattering processes with several particles in the final state.
}

\myKeywords{QCD}

%

\section{Introduction\label{Sec:intro}}
Besides determining the precise nature of the recently discovered Higgs-like boson, the scientific plans of the Large Hadron Collider (LHC) include further tests of complex dynamics of the Standard Model and searches of physics beyond the Standard Model \cite{Salgado:2011wc}.
%
Such a scientific program requires also theoretical control over background and signal processes occuring at the collisions.
A substantial part of the necessary calculations involves the application of Quantum Chromodynamics (QCD).
It enables calculations for LHC physics via various factorization theorems, allowing for the decomposition of a given process into a part characterizing the colliding hadrons, called the parton densities, and a so-called hard part characterizing the parton parton scattering and production process.
The predictive power of QCD largely relies on the fact that this hard part can be calculated within perturbation theory.
%
%
%
Regarding the factorization theorems, we will focus on high-energy factorization~\cite{Catani:1994sq} here, which applies when the energy scale involved in the scattering process is high and larger then any other scale in the process.
For recent applications see~\cite{Hautmann:2012rf,Hautmann:2012qr,Hautmann:2012sh,Kutak:2012rf,Deak:2011ga,Deak:2010gk,Deak:2009xt}.
This approach is particularly interesting as far as LHC physics is concerned since the energy is indeed the largest available scale and therefore this approach should provide the optimal method to calculate cross sections for various processes.
Furthermore, high-energy factorization already at lowest perturbatively defined order provides kinematical effects in hard matrix elements which are relegated in other approaches to corrections of higher orders.
The evolution equations of high energy factorization providing transversal momentum dependent parton densities sum up logarithms of energy accompanied by a coupling constant~\cite{Kovchegov:1999yj,Balitsky:1995ub,Kutak:2011fu,Kutak:2012qk}.
Because of the dependence on the parton's transversal momentum, parton densities of high-energy factorization have to be convoluted with the hard process which is calculated with the initiating gluons being off-shell.

Several frameworks exist to calculate gauge invariant matrix elements for partonic processes initiated by off-shell gluons~\cite{Lipatov:1995pn,Antonov:2004hh,DelDuca:1999ha,Andersen:2009nu}.
%
In order for such a framework to be widely used, a practical numerical tool would be desirable, like the ones that already exist for the calculation of tree-level on-shell scattering amplitudes, \eg~\cite{Mangano:2002ea,Cafarella:2007pc,Gleisberg:2008fv,Kleiss:2010hy,Alwall:2011uj}.
They operate on the amplitude level and use helicity methods, and thus are very efficient and universal, and can deal with essentially an arbitrary number of external gluons.
The present study is a generalization of~\cite{vanHameren:2012uj}, and provides the proof of concept for the application of the helicity method to high-energy factorization with two initial state gluons being off-shell.
It includes an implementation of this method in a form of a numerical tool.
In particular we develop a new prescription for calculating gauge invariant high-energy factorizable amplitudes for any tree-level process with two gluons in the initial state being off-shell.
Our results turn out to be in agreement with results obtained using Lipatov's action.

In short, our prescription is established as follows.
It is known~\cite{Leonidov:1999nc} that the usual Feynman graphs contributing to an amplitude in an on-shell calculation are in general not sufficient to obtain gauge invariant results if any of the external legs are off-shell.
Thus the off-shell process needs to be embedded into a larger on-shell process, from which a gauge invariant off-shell amplitude needs to be disentangled.
%
The first crucial point of our approach is that we manage to have exactly the kinematics of high-energy factorization for the off-shell process, and at the same time have exact on-shellness for the embedding, through the application of complex momenta~\cite{Britto:2005fq} for the on-shell ``external'' partons of the embedding.
%
The second crucial point is that we have a well-defined prescription how to decouple the unneeded on-shell partons.

The paper is organized as follows.
In the \Section{Sec:prescription} we introduce the kinematics of high-energy factorization and introduce our prescription which provides gauge invariant amplitudes for any tree-level process with off-shell initial-state gluons.
In \Section{Sec:derivation} we provide the derivation of our prescription using the spinor helicity formalism and analytic continuation of the amplitude to a region where the momenta of the auxiliary partons are complex.
%
%
Furthermore, we show that the obtained amplitudes obey the correct collinear limit when the initial-state gluons become on-shell.
In \Section{Sec:OneOffShell} we discuss the connection of our method with our former result~\cite{vanHameren:2012uj} for $g^*g\to N g$.
In \Section{Sec:Lipatov} we show that our prescription reproduces the results obtained from the effective action approach, \ie\ that the effective vertices of~\cite{Antonov:2004hh} correspond to our amplitudes with reggeized gluons.
In \Section{Sec:results} we present the numerical application of our framework to various LHC processes with a simple model for the transversal-momentum dependent parton density function.
We conclude in \Section{Sec:conclusion}, and the appendices contain details of the spinor formalism and explicit calculations of reggeon-gluon-gluon and reggeon-reggeon-gluon vertices.
%

\section{Prescription\label{Sec:prescription}}
We start with stating the main result of this paper, which is a prescription to calculate scattering amplitudes in quasi-multi-Regge kinematics in the high-energy limit, \ie\ scattering amplitudes with off-shell initial-state gluons, and with no kinematical restrictions on the final-state particles other than momentum conservation and on-shellness.
The derivation will be given in the next section.

The exact definition of the scattering amplitudes we deal with requires the introduction of two momenta $\ell_1,\ell_2$ such that
%
\begin{equation}
\lid{\ell_1}{\ell_1} = \lid{\ell_2}{\ell_2} = 0
\quad,\quad
\lid{\ell_1}{\ell_2} = 2E^2
\quad,\quad
E>0
~.
\end{equation}
%
The high-energy limit is enforced by demanding that the momenta of the off-shell initial-state gluons are given by
%
\begin{equation}
k_1^\mu = \xone \ell_1^\mu + k_{1\perp}^\mu
\quad,\quad
k_2^\mu = \xtwo \ell_2^\mu + k_{2\perp}^\mu
\label{eqn:k1k2}
\end{equation}
%
for $\xone,\xtwo\in[0,1]$ and
%
\begin{equation}
 \lid{\ell_1}{k_{1\perp}}
=\lid{\ell_2}{k_{1\perp}}
=\lid{\ell_1}{k_{2\perp}}
=\lid{\ell_2}{k_{2\perp}}
=0
~.
\end{equation}
%
The expressions for the off-shell momenta are restricted in the sense that for $k_1^\mu$ a component proportional to $\ell_2^\mu$ is missing, and for $k_2^\mu$ a component proportional to $\ell_1^\mu$ is missing.
This is the high-energy limit, or high-energy approximation.
In our derivation, however, it will not be the result of an approximation.
The kinematics of \Equation{eqn:k1k2} is set from the start.
The prescription to calculate the scattering amplitude for the process
%
\begin{equation}
g^*(k_1)\,g^*(k_2)\,\to\,X(k_1+k_2=p_1+p_2+\ldots+p_n)
\nonumber
\end{equation}
%
with $n$ particles in the final state is to embed the process into a quark-scattering process, and apply eikonal Feynman rules to the quark lines.
More precisely:
\begin{enumerate}
\item
\raisebox{-44.5pt}{
\parbox{0.71\linewidth}{
Consider the process $q_A\,q_B\to q_A\,q_B\,X$, where $q_A,q_B$ are distinguishable massless quarks not occurring in $X$, and with momentum flow as if the momenta $p_A,p_B$ of the initial-state quarks and $p_{A'},p_{B'}$ of the final-state quarks are given by
\begin{equation}
p_A^\mu=k_1^\mu
\quad,\quad
p_B^\mu=k_2^\mu
\quad,\quad
p_{A'}^\mu=p_{B'}^\mu=0
~.
\nonumber
\end{equation}
}
\hspace{\fill}
\parbox{0.28\linewidth}{
\epsfig{figure=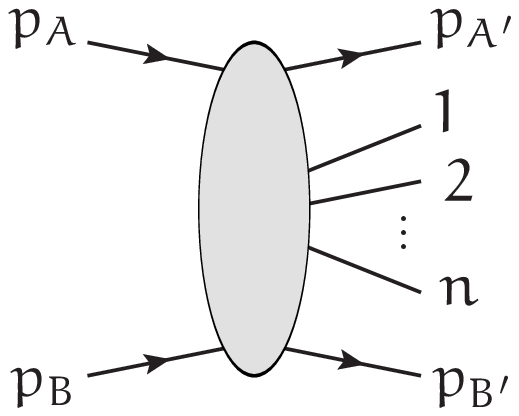,width=\linewidth}
}
}
\item Associate the number $1$ instead of spinors with the end points of the $A$-quark line,\\
      interpret every vertex on the $A$-quark line as $\gqcd T^a_{ij}\,\ell_1^\mu$ instead of $-\imag\gqcd T^a_{ij}\,\gamma^\mu$,\\
      interpret every propagator on the $A$-quark line as $\delta_{ij}/\lid{\ell_1}{p}$ instead of $\imag\delta_{ij}/\slashp$.
\item Associate the number $1$ instead of spinors with the end points of the $B$-quark line,\\
      interpret every vertex on the $B$-quark line as $\gqcd T^a_{ij}\,\ell_2^\mu$ instead of $-\imag\gqcd T^a_{ij}\,\gamma^\mu$,\\
      interpret every propagator on the $B$-quark line as $\delta_{ij}/\lid{\ell_2}{p}$ instead of $\imag\delta_{ij}/\slashp$.
\item Multiply the amplitude with
\begin{equation}
F =
\frac{\imag\,\xone\sqrt{-2k_{1\perp}^2}}{\gqcd}
\times
\frac{\imag\,\xtwo\sqrt{-2k_{2\perp}^2}}{\gqcd}
~.
\label{eqn:factor}
\end{equation}
\end{enumerate}
For the rest, normal Feynman rules apply.
The coupling constant in the denominators of (\ref{eqn:factor}) are obviously necessary because the quark-scattering process carries a higher power of the coupling constant.
The factor includes $\sqrt{2}\times\sqrt{2}$ to correct for the difference in color representation between the quarks and gluons, and a factor $\imag\xone\sqrt{-k_1^2}\times\imag\xtwo\sqrt{-k_2^2}$ is necessary to obtain the correct collinear limit.
%

\section{Derivation\label{Sec:derivation}}
The prescription involves embedding the process with the off-shell initial-state gluons into the quark-scattering process $q_A\,q_B\to q_A\,q_B\,X$, with eikonal Feynman rules for the quark-lines, and deformed kinematics.
The starting point of the derivation will be the scattering amplitude for the on-shell process, with normal quark lines
%
\begin{equation}
\Amp(q_A\,q_B\to q_A\,q_B\,X)
~.
\label{eqn:bedAmp}
\end{equation}
%
This is a well-defined gauge invariant function of the external momenta, helicities and colors.
The key-point of our derivation is that we will not apply any approximations througout, and will manifestly keep gauge invariance throughout.
The high-energy limit of \Equation{eqn:k1k2} will be set from the start.
We do need, however, to extrapolate the amplitude~\Ref{eqn:bedAmp} beyond its physical interpretation.
More explicitly, we will use the fact that in order to guarantee gauge invariance of~\Ref{eqn:bedAmp} the external momenta need to be on-shell and need to satisfy momentum conservation, but are not required to be real (\cf\ \cite{Britto:2005fq}).
%

\subsection{Gauge invariant helicity amplitudes}
Rather than putting everything in an appendix, we state some essential points regarding the formalism here.
Consider the following decomposition of four-momenta \cite{delAguila:2004nf}.
Given two light-like momenta $\ell_1,\ell_2$ such that $\lid{\ell_1}{\ell_2}\neq0$, we define
%
%
\begin{equation}
\ell_3^\mu=\srac{1}{2}\bram{\ell_2}\,\gamma^\mu\,\ketm{\ell_1}
\quad,\quad
\ell_4^\mu=\srac{1}{2}\bram{\ell_1}\,\gamma^\mu\,\ketm{\ell_2}
~.
\label{eqn:l3l4}
\end{equation}
%
The exact definition of the spinors in the expressions above are given in \Appendix{App:spinors}.
We have
%
\begin{equation}
 \lid{\ell_1}{\ell_1}
=\lid{\ell_2}{\ell_2}
=\lid{\ell_3}{\ell_3}
=\lid{\ell_4}{\ell_4}
=\lid{\ell_1}{\ell_3}
=\lid{\ell_1}{\ell_4}
=\lid{\ell_2}{\ell_3}
=\lid{\ell_2}{\ell_4}
=0
~,
\end{equation}
%
while
%
\begin{equation}
\lid{\ell_3}{\ell_4} = -\lid{\ell_1}{\ell_2}
~.
\end{equation}
%
Any four-vector $p$ can now be decomposed as
%
\begin{equation}
p^\mu = \frac{\lid{\ell_2}{p}}{\lid{\ell_1}{\ell_2}}\,\ell_1^\mu
      + \frac{\lid{\ell_1}{p}}{\lid{\ell_1}{\ell_2}}\,\ell_2^\mu
      - \frac{\lid{\ell_4}{p}}{\lid{\ell_1}{\ell_2}}\,\ell_3^\mu
      - \frac{\lid{\ell_3}{p}}{\lid{\ell_1}{\ell_2}}\,\ell_4^\mu
~.
\end{equation}
%
The first two terms are the usual terms in the Sudakov parametrization in terms of $\ell_1,\ell_2$, and the last two terms form an explicit decomposition of the transversal component.
If $\ell_1,\ell_2$ are real, then the momenta $\ell_3,\ell_4$ are in general complex.
%
The spinors in \Appendix{App:spinors} are well-defined also for complex momenta, and we have the following identities
%
\begin{align}
\bram{\ell_3} = \bram{\ell_2} \quad&,\quad \ketm{\ell_3} = \ketm{\ell_1}
\notag\\
\brap{\ell_3} = \brap{\ell_1} \quad&,\quad \ketp{\ell_3} = \ketp{\ell_2}
\notag\\
\bram{\ell_4} = \bram{\ell_1} \quad&,\quad \ketm{\ell_4} = \ketm{\ell_2}
\label{eqn:l3l4spin}\\
\brap{\ell_4} = \brap{\ell_2} \quad&,\quad \ketp{\ell_4} = \ketp{\ell_1}
\notag
~.
\end{align}

The amplitude of the processs $q_A\,q_B\to q_A\,q_B\,X$ contains Feynman graphs that consist of a quark-line of type $A$ with a single gluon attached and a quark-line of type $B$ with a single gluon attached.
They are represented by the first term on the r.h.s.\ in \Figure{Fig1}.
\myFigure{
\epsfig{figure=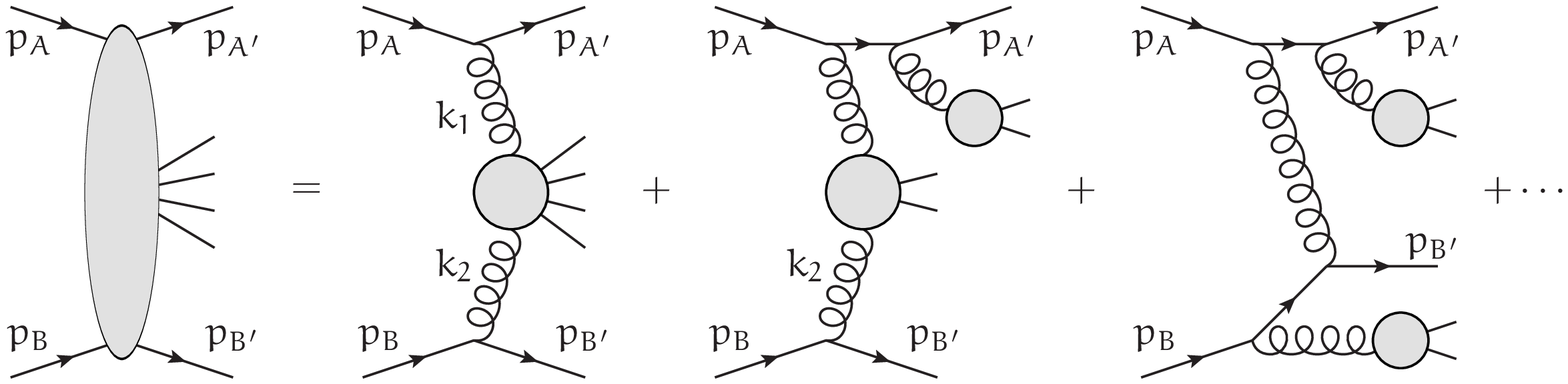,width=0.94\linewidth}
\caption{\label{Fig1}A few terms in the classification of the graphs contributing to $q_A\,q_B\to q_A\,q_B\,X$ w.r.t.\ the gluons attached to the quark lines.}
}
The heuristic picture is that these gluons are going to play the role of the off-shell gluons in $g^*g^*\to X$.
It needs to be stressed that we will not drop any of the other graphs and that we keep the full gauge invariant sum.
We demand that 
%
\begin{equation}
p_A^\mu-p_{A'}^\mu=k_1^\mu=\xone\ell_1^\mu+k_{1\perp}^\mu
\quad,\quad
p_B^\mu-p_{B'}^\mu=k_2^\mu=\xtwo\ell_2^\mu+k_{2\perp}^\mu
\label{eqn:pApAp}
~.
\end{equation}
%
We do not set $p_A$ equal to $\ell_1$.
We are interested in obtaining a scattering amplitude for the process~$g^*g^*\to X$, and are not concerned about the momenta of the quarks.
We only need to make sure we have overall momentum conservation and on-shellness.
At first sight, the latter requirement seems to be in contradiction with requirements~\Ref{eqn:pApAp}, and it would be if we required the momenta of the quarks to be real.
Choosing, however,
%
\begin{align}
p_A^\mu = (\Lambda+\xone)\ell_1^\mu + \kappa_1^3\ell_3^\mu
\quad&,\quad
p_{A'}^\mu = \Lambda\ell_1^\mu - \kappa_1^4\ell_4^\mu
\notag\\
p_B^\mu = (\Lambda+\xtwo)\ell_2^\mu + \kappa_2^4\ell_4^\mu
\quad&,\quad
p_{B'}^\mu = \Lambda\ell_2^\mu - \kappa_2^3\ell_3^\mu
\label{eqn:pApB}
\end{align}
%
with
%
\begin{align}
\kappa_1^{3}=-\frac{\lid{\ell_4}{k_{1\perp}}}{\lid{\ell_1}{\ell_2}}
\quad&,\quad
\kappa_1^{4}=-\frac{\lid{\ell_3}{k_{1\perp}}}{\lid{\ell_1}{\ell_2}}
\notag\\
\kappa_2^{3}=-\frac{\lid{\ell_4}{k_{2\perp}}}{\lid{\ell_1}{\ell_2}}
\quad&,\quad
\kappa_2^{4}=-\frac{\lid{\ell_3}{k_{2\perp}}}{\lid{\ell_1}{\ell_2}}
~,
\label{eqn:defKappa}
\end{align}
%
both on-shellness of $p_A,p_{A'},p_B,p_{B'}$ and \Equation{eqn:pApAp} are assured to hold, for any value of $\Lambda$.
Furthermore, following the identities~\Ref{eqn:l3l4spin}, we have
%
\begin{align}
\bram{p_{A'}}\propto\bram{\ell_1} \quad&,\quad \ketm{p_A}\propto\ketm{\ell_1}
\notag\\
\bram{p_{B'}}\propto\bram{\ell_2} \quad&,\quad \ketm{p_B}\propto\ketm{\ell_2}
\label{eqn:pABspin}
~,
\end{align}
%
so we may simply assign the following spinors to the quarks in the amplitude without spoiling gauge invariance:
%
\begin{align}
q_A(p_A)\to\ketm{\ell_1} \quad&,\quad q_A(p_{A'})\to\bram{\ell_1}
\notag\\
q_B(p_B)\to\ketm{\ell_2} \quad&,\quad q_B(p_{B'})\to\bram{\ell_2}
~.
\label{eqn:qAqB}
\end{align}
%
As a result, in the graphs of the first term on the r.h.s.\ of \Figure{Fig1}, both quark-lines are connected to gluons with eikonal vertices
%
\begin{equation}
\bram{\ell_1}\,\gamma^\mu\,\ketm{\ell_1} = 2\ell_1^\mu
\quad,\quad
\bram{\ell_2}\,\gamma^\mu\,\ketm{\ell_2} = 2\ell_2^\mu
~.
\end{equation}
%
Other graphs contain $A$-quark propagators and/or $B$-quark propagators, and depend on $\Lambda$.
We want to stress that the constructed amplitude is gauge invariant {\em for any value of $\Lambda$}.

Now, we consider the limit
%
\begin{equation}
\Lambda\to\infty
~,
\label{eqn:limit}
\end{equation}
%
and we will see that it is a well-defined limit.
Considering the role of $\Lambda$ in \Equation{eqn:pApB}, one might interprete it as the high-energy limit.
Realize, however, that $\Lambda$ is a dimensionless parameter, not the energy, and that the limit is not an approximation.
The actual high-energy approximation was essentially made by requiring \Equation{eqn:k1k2} from the start.
The limit~\Ref{eqn:limit} should rather be seen as the final step to extract a physical amplitude, independent of imaginary momentum components.
The momentum $p$ of an $A$-quark propagator can be written as
%
\begin{equation}
p^\mu = (\Lambda+x_p)\ell_1^\mu+y_p\ell_2^\mu+p_\perp^\mu
~,
\end{equation}
%
for some numbers $x_p,y_p$ and (complex) $p_\perp$, which are independent of $\Lambda$.
Since $\lid{\ell_1}{p_{A'}}=0$, we can write
%
\begin{equation}
y_p\,\lid{\ell_1}{\ell_2} = \lid{\ell_1}{p}
= \lid{\ell_1}{(p-p_{A'})}
= \lid{\ell_1}{p'}
~,
\end{equation}
%
where $p'=p-p_{A'}$ is the momentum that would be flowing in the propagator if the external quarks had the momenta $p_A=k_1$ and $p_{A'}=0$.
The square of the momentum is given by
%
\begin{equation}
p^2 = 2(\Lambda+x_p)\,y_p\,\lid{\ell_1}{\ell_2} + p_\perp^2
~,
\end{equation}
%
so for the propagator of an $A$-quark line with momentum $p$ we find
%
\begin{equation}
\frac{\slashp}{p^2}
\;=\;
\frac{(\Lambda+x_p)\slashl_1+y_p\slashl_2+\slashp_\perp}
     {2(\Lambda+x_p)\,y_p\,\lid{\ell_1}{\ell_2} + p_\perp^2}
\;\overset{\Lambda\to\infty}{\longrightarrow}\;
\frac{\slashl_1}{2\,y_p\,\lid{\ell_1}{\ell_2}}
=
\frac{\slashl_1}{2\,\lid{\ell_1}{p'}}
~.
\label{eqn:eikDen}
\end{equation}
%
%
Similarly, the propagator of a $B$-quark line with momentum $p$ will become $\slashl_2/(2\,\lid{\ell_2}{p'})$, where $p'=p-p_{B'}$ is the momentum that would be flowing in the propagator if the external quarks had the momenta $p_B=k_2$ and $p_{B'}=0$.
Any gluon attached to an $A$-or $B$-quark line will do so via an eikonal vertex, \eg\ for the $A$-quark line we have
%
\begin{align}
 \bram{\ell_1}\,\gamma^{\mu_1}\,\slashl_1\,\gamma^{\mu_2}\,\slashl_1\,\cdots\,\ketm{\ell_1}
&=\bram{\ell_1}\,\gamma^{\mu_1}\,\ketm{\ell_1}\bram{\ell_1}\,\gamma^{\mu_2}\,\ketm{\ell_1}\bram{\ell_1}\,\cdots\,\ketm{\ell_1}
\notag\\
&= (2\ell_1^{\mu_1})(2\ell_1^{\mu_2})\cdots
~.
\label{eqn:eikNum}
\end{align}
%
The factors $(-2\imag)$ from the vertices and $(\imag/2)$ from the propagators on the eikonal lines cancel, except for one vertex on each line.
This happens in each graph, and thus leads to an overall factor $2\times2$, which we remove.
This concludes the derivation of the prescription up to the last point about the extra factor.

\subsection{Matching to the collinear limit}
Let us denote the obtained amplitude by
%
\begin{equation}
\Amp^{(\infty)}_{i_Aj_A,i_Bj_B}(q_A\,q_B\to q_A\,q_B\,X)
~.
\label{eqn:Ahef}
\end{equation}
%
The sub-scripts indicate the color degrees of freedom of the quarks.
As explained earlier, the coupling constant in the denominators is obviously necessary in \Equation{eqn:factor}.
We will now show that the amplitude also has to be multiplied with the numerator.
In order to arrive at a satisfactory helicity amplitude for the process $g^*g^*\to X$, the obtained amplitude needs to be matched at the collinear limit of $k_{1\perp}^2,k_{2\perp}^2\to0$ to the process $g\,g\to X$ with on-shell initial-state gluons.
We perform the matching at the level of the squared amplitude, averaged over initial-state helicities for the on-shell process, and averaged over the azimuthal angle of the transversal components of the off-shell gluon momenta for the off-shell process~\cite{Catani:1990eg}.
The color content of the obtained amplitude is different from the content of the on-shell process, and the matching of these requires also summation over initial-state colors.
%
So the factor $F$ in \Equation{eqn:factor} should be such that
%
\begin{multline}
\frac{1}{4}\,\sum_{a_1,a_2}\sum_{\lambda_1,\lambda_2}
\left|\Amp^{a_1a_2}_{\lambda_1\lambda_2}(g\,g\to X)\right|^2
=\\
\sum_{i_A,j_A,i_B,j_B}
\left\langle\left|\lim_{k_{1\perp}^2,k_{2\perp}^2\to0}F\,\Amp^{(\infty)}_{i_Aj_A,i_Bj_B}(q_A\,q_B\to q_A\,q_B\,X)\right|^2\right\rangle_{\phi_1,\phi_2}
~.
\end{multline}
%
The average over the azimuthal angles can be defined \eg\ for $k_{1\perp}$ by parameterizing it in terms of $|k_{1\perp}|$ and $\phi_1$ following
%
\begin{equation}
k_{1\perp}^\mu = \frac{|k_{1\perp}|}{\sqrt{2\lid{\ell_1}{\ell_2}}}
             \left(e^{\imag\phi_1}\ell_3^\mu + e^{-\imag\phi_1}\ell_4^\mu\right)
~,
\end{equation}
%
and performing the integral over $\phi_1$, dividing the result by $2\pi$.

\subsubsection{Color\label{Sec:color}}

The off-shell gluons in the process $g^*g^*\to X$ carry color degrees of freedom in the adjoint representation.
The additional quarks in the process $q_A\,q_B\to q_A\,q_B\,X$ are in the fundamental representation.
Eventually, the amplitude needs to be squared and summed over color degrees of freedom.
Abreviating $\Amp^{a,b}(g^*g^*\to X)$ just to $\Amp^{a,b}$, where $a,b$ are the color indices of the off-shell gluons, we have
%
\begin{align}
\sum_{a,b}\big|\Amp^{a,b}\big|^2
&=
\sum_{a,b,c,d}\Amp^{a,b}\,\delta^{ac}\delta^{bd}\,\big(\Amp^{c,d}\big)^*
\notag\\
&=
\sum_{a,b,c,d}\Amp^{a,b}\big(2\Tr T^aT^c\big)\big(2\Tr T^bT^d\big)\big(\Amp^{c,d}\big)^*
=
\sum_{i,j,k,l}\big|2\Amp'_{ij,kl}\big|^2
~,
\end{align}
%
where
%
\begin{equation}
\Amp'_{ij,kl}(g^*g^*\to X)
= \sum_{a,b}T^{a}_{ij}T^b_{kl}\,\Amp^{a,b}(g^*g^*\to X)
~.
\end{equation}
%
So we see that attaching the quark lines to the off-shell gluons does not alter the color content of the amplitude apart from a factor $2$ if, eventually, we sum over the color degrees of freedom of the quarks.
As an alternative approach, we could simply interpret the amplitude to be in the color-flow representation \cite{Kanaki:2000ms,Maltoni:2002mq}, which would also require an extra factor $\sqrt{2}\times\sqrt{2}$.
This explains the factor $\sqrt{2}\times\sqrt{2}$ in \Equation{eqn:factor}.

\subsubsection{Collinear limit\label{Sec:collinear}}
We consider the amplitude of \Equation{eqn:Ahef}.
\myFigure{%
\epsfig{figure=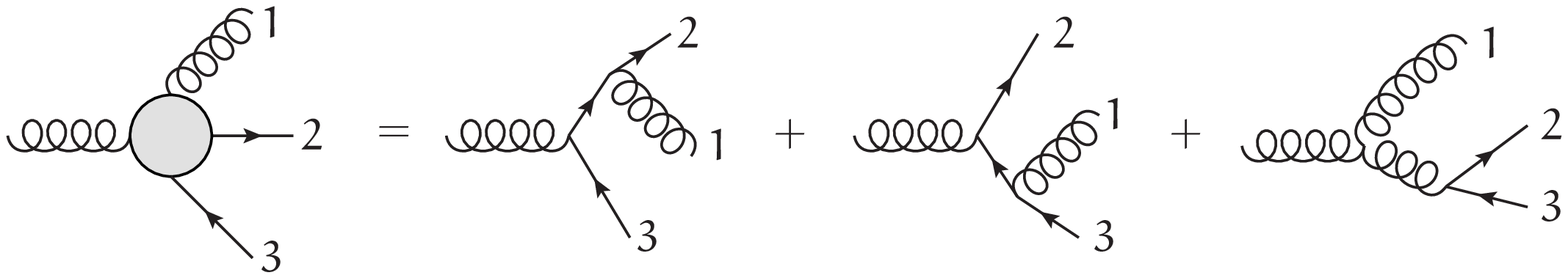,width=0.9\linewidth}
\caption{\label{Fig2}An example of an off-shell current. The enumerated external lines are on-shell, the other gluon is off-shell with momentum $k=p_1+p_2+p_3$.}
}
Let $J^a_\mu$ be an off-shell current, \ie\ the sum of all connected sub-graphs that contain a given set of the external particles plus one off-shell gluon.
An example is depicted in \Figure{Fig2}.
The propagator of the off-shell gluon is not included in the off-shell current.
The momentum $k^\mu$ of this gluon is the sum of the momenta of the external particles in $J^a_\mu$.
Gauge invariance implies current conservation, which dictates that
%
\begin{equation}
 k^\mu\,J^a_\mu = 0
~.
\end{equation}
%
If this off-shell current is attached to the $A$-quark line via the numerator of a gluon propagator $d^\nu_\mu(k,n)$ in an axial gauge with gauge vector $n^\mu$, we have
%
\begin{align}
\ell_1^\mu\,d^\nu_\mu(k,n)\,J^a_\nu
&=\ell_1^\mu\left(-g_\mu^\nu+\frac{k_\mu n^\nu+n_\mu k^\nu}{\lid{n}{k}}-n^2\,\frac{k_\mu k^\nu}{(\lid{n}{k})^2}\right)J^a_\nu
\notag\\
&=-\big(\ell_1^\mu J^a_\mu\big) + \frac{\lid{\ell_1}{k}}{\lid{n}{k}}\,\big(n^\mu J^a_\mu\big)
~.
\label{eqn:ALine}
\end{align}
%
For the gluon with momentum $k_1^\mu=\xone\ell_1^\mu+k_{1\perp}^\mu$, we have $\lid{\ell_1}{k_1}=0$, and we see that the corresponding off-shell current $J^a_{1\mu}$ is always attached to the $A$-quark line as $\big(\!-\ell_1^\mu J^a_{1\mu}\big)$, independently from the gauge.
Using current conservation again, $k_1^\mu J^a_{1\mu}=0$, we have
%
\begin{equation}
-\ell_1^\mu J^a_{1\mu} = \frac{1}{\xone}\,k_{1\perp}^\mu J^a_{1\mu}
~.
\end{equation}
%
Including the remaining piece of propagator $\imag/k_{1\perp}^2$, we see that
the off-shell gluon with momentum $k_1$ is attached to the $A$-quark line via the contraction with
%
\begin{equation}
\frac{-\imag}{\xone\sqrt{-k_{1\perp}^2}}\times\frac{k_{1\perp}^\mu}{\sqrt{-k_{1\perp}^2}}\,
~.
\label{eqn:perp1}
\end{equation}
%
The second part in the expression above will give the correct average over the azimuthal angle, and the first part represents the reciprocal contribution to the factor of \Equation{eqn:factor}.
Similarly, for the gluon with momentum $k_2^\mu=\xtwo\ell_2^\mu+k_{2\perp}^\mu$ we find that it is attached to the $B$-quark line via the contraction with 
%
\begin{equation}
\frac{-\imag}{\xtwo\sqrt{-k_{2\perp}^2}}\times\frac{k_{2\perp}^\mu}{\sqrt{-k_{2\perp}^2}}\,
~.
\label{eqn:perp2}
\end{equation}
%

For $k_{1\perp}^2,k_{2\perp}^2\to0$, contributions that do not contain both off-shell gluons with the momenta $k_1,k_2$ vanish, and only those from the first term on the r.h.s.\ of \Figure{Fig1} remain, which are exactly the ones one would take into account for the on-shell process $gg\to X$.
The only difference is that these gluons are contracted with~\Ref{eqn:perp1} and~\Ref{eqn:perp2} instead of polarization vectors.
We consider the collinear limit at the level of the squared amplitude, averaged over the azimuthal angles of $k_{1\perp}$ and $k_{2\perp}$.
Notice that for $k_{1\perp}^2,k_{2\perp}^2\to0$, the amplitude only depends on $k_{1\perp},k_{2\perp}$ through the contractions with~\Ref{eqn:perp1} and~\Ref{eqn:perp2}, and it is enough to prove that the contractions of the squared amplitude with
%
\begin{equation}
\bigg\langle\frac{k_{1\perp}^\mu k_{1\perp}^\nu}{-k_{1\perp}^2}\bigg\rangle_{\phi}
\quad,\quad
\bigg\langle\frac{k_{2\perp}^\mu k_{2\perp}^\nu}{-k_{2\perp}^2}\bigg\rangle_{\phi}
\end{equation}
%
lead to the correct average over initial-state polarizations.
We just consider $k_{1\perp}$, the other one goes analogously.
Using its decomposition~\Ref{eqn:defKappa} in terms of $\ell_3,\ell_4$, we have
%
\begin{equation}
\frac{k_{1\perp}^\mu k_{1\perp}^\nu}{-k_{1\perp}^2}
=
 \frac{\kappa_1^3}{\kappa_1^4}\frac{\ell_3^\mu\ell_3^\nu}{2\lid{\ell_1}{\ell_2}}
+\frac{\kappa_1^4}{\kappa_1^3}\frac{\ell_4^\mu\ell_4^\nu}{2\lid{\ell_1}{\ell_2}}
+\frac{\ell_3^\mu\ell_4^\nu+\ell_4^\mu\ell_3^\nu}{2\lid{\ell_1}{\ell_2}}
~.
\label{eqn:polsum}
\end{equation}
%
Rembering \Equation{eqn:l3l4}, we see that $\ell_3^\mu/\sqrt{\lid{\ell_1}{\ell_2}}$ and $\ell_4^\nu/\sqrt{\lid{\ell_1}{\ell_2}}$ are just the two polarization vectors in the Kleiss-Stirling \cite{Kleiss:1985yh} construction for light-like momentum $\ell_1$ with reference momentum $\ell_2$, and that the third term on the r.h.s.\ in \Equation{eqn:polsum} gives the desired polarization average.
The dependence of $\kappa_1^3/\kappa_1^4$ on $\phi$ is proportional to $e^{2\imag\phi}$, so that the first two terms vanish in the average over $\phi$.

\section{\label{Sec:OneOffShell}Processes with one off-shell gluon}

For processes with only one off-shell gluon, \ie\ $g^*\,y\to X$, with $y=g,q,\overline{q}$, one has to consider the process $q_A\,y\to q_A\,X$, and only apply the eikonal Feynman rules to the $A$-quark line.
In this case, the helicity amplitude can be simplified further by using a special axial gauge.

In the previous section we saw already that the off-shell current $J^a_{1\mu}$ with momentum $k_1=\xone\ell_1+k_{1\perp}$ is attached to the $A$-quark line as $\big(\!-\ell_1^\mu J^a_{1\mu}\big)$, independently from the gauge.
\Equation{eqn:ALine} also tells us that {\em if we choose the axial gauge with $n^\mu=\ell_1^\mu$, then all contributions to the amplitude with any other off-shell current attached to the $A$-quark line vanish}.
In other words, the only graphs with gluons attached to the $A$-quark line that need to be taken into account are one of the following:
\begin{enumerate}
\item the only gluon attached to the $A$-quark line is the gluon with momentum $k_1^\mu=\xone\ell_1^\mu+k_{1\perp}^\mu$
\item all gluons attached to the $A$-quark line are on-shell
\end{enumerate}
Consequently, if $X$ in the process $g^*\,y\to X$ does not consist of gluons only, the second option above cannot occur, and we see that
\begin{itemize}
\item for the process $g^*\,y\to X$ where $X$ does not consist of gluons only, helicity amplitudes can be calculated by applying Feynman rules without attaching a propagator, in the axial gauge with gauge vector $\ell_1$, and by contracting the off-shell gluon with
\begin{equation}
\frac{k_{1\perp}^\mu}{\sqrt{-k_{1\perp}^2}}
~.
\end{equation}
\end{itemize}
We assume the off-shell gluon to be in the adjoint color representation here.
If $X$ consists of gluons only, then the contribution coming from graphs with all on-shell gluons attached to the $A$-quark line is exactly the `gauge-restoring' amplitude of equation~(25) in~\cite{vanHameren:2012uj}.
It can be eliminated by choosing the reference momentum of the polarization vector $\varepsilon^\mu$ of at least one of the on-shell gluons equal to $\ell_1^\mu$, so that for this on-shell gluon $\lid{\ell_1}{\varepsilon}=0$.

\section{\label{Sec:Lipatov}Effective reggeon vertices}
In this section we will show that the prescription for the calculation of scattering amplitudes with off-shell gluons can reproduce the results of~\cite{Antonov:2004hh}.
The effective vertices involving reggeized gluons presented in that paper correspond to amplitudes with off-shell gluons as presented here, in the case that the non-reggeized gluons are on-shell.
Reggeized gluons in these vertices correspond to gluons with momenta restricted to be of the form of \Equation{eqn:k1k2}.
To be more specific, in the following, we show that our prescription provides
%
\begin{equation}
\frac{-\xone E}{\sqrt{-k_1^2}}\,\sqrt{2}T^c_{ij}\times
\frac{-\xtwo E}{\sqrt{-k_2^2}}\,\sqrt{2}T^d_{kl}\times
\Gamma^{-\mu_1\mu_2\cdots\mu_n+}_{ca_1a_2\cdots a_nd}
       (k_1;p_1,p_2,\ldots,p_n;k_2)
~,
\end{equation}
%
where $\Gamma$ is the reggeon-reggeon-$n$-particle vertex from~\cite{Antonov:2004hh}.
Analogously, the prescription provides the single-reggeon vertices.

\myFigure{%
\epsfig{figure=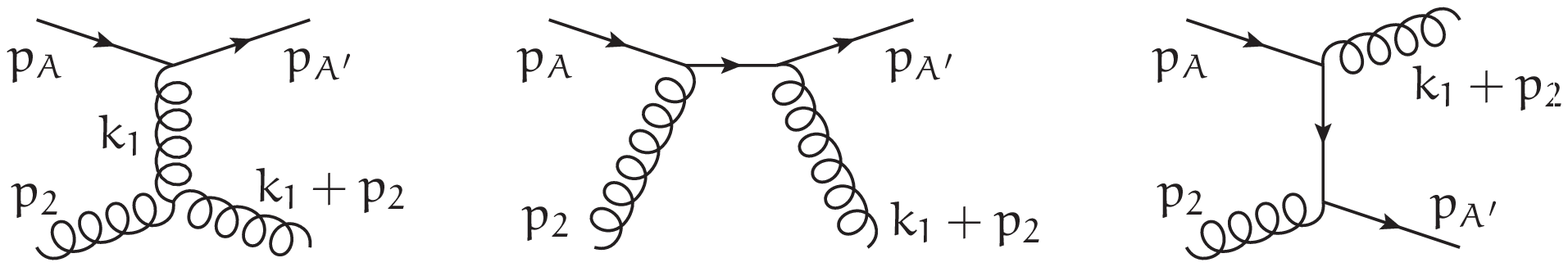,width=0.7\linewidth}
\caption{\label{Fig3}Feynman graphs for $q_A\,g\to q_A\,g$.}
}
 
\myFigure{%
\epsfig{figure=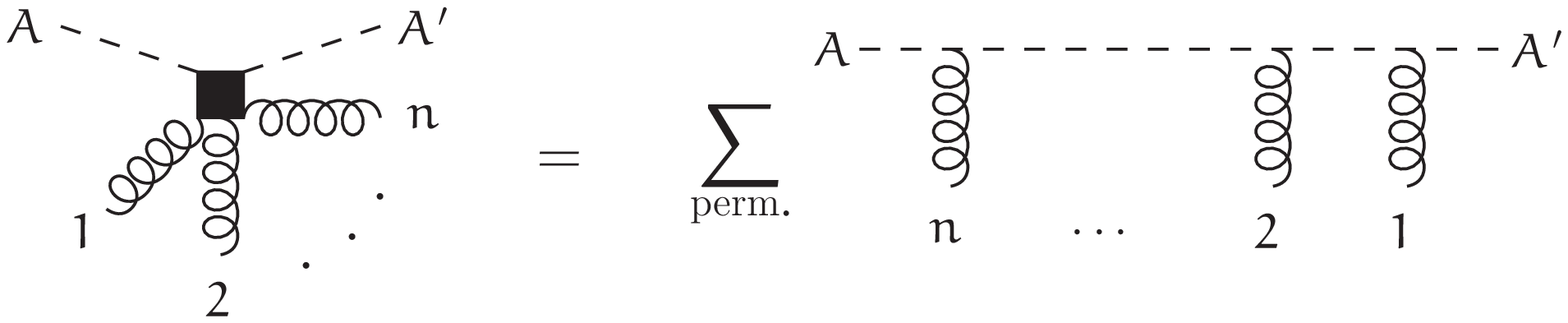,width=0.70\linewidth}
\caption{\label{Fig4}Effective vertices consisting of $n$ gluons attached to an eikonal line. The sum is over all permutations of $(1,2,\ldots,n)$.}
}
 
We present two explicit calculations in \Appendix{App:Lipatov}, namely for $g^*g\to g$ and $g^*g^*\to g$.
The necessary graphs in the embedding of $g^*g\to g$ are depicted in \Figure{Fig3}.
The last two graphs in \Figure{Fig3} contribute as an effective vertex of the type of \Figure{Fig4}.
These vertices contain all possible ways a number of gluons can be attached directly to the eikonal line.
According to our prescription, $p_{A'}=0$, and the denominator factor of each propagator is given by $-\lid{\ell_1}{p}$ where $p$ is the sum of the incoming momenta of the gluons to the right of the propagator.
We have
%
\begin{equation}
\raisebox{-28pt}{\epsfig{figure=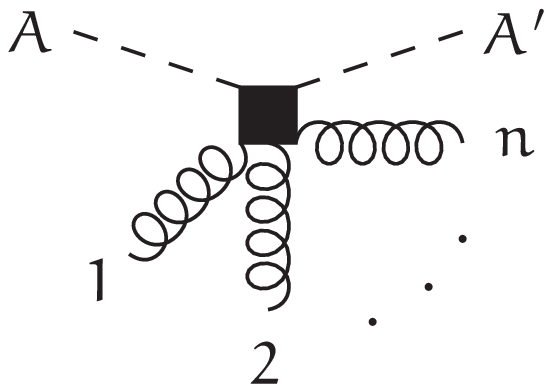,width=0.2\linewidth}}
=
\imag\gqcd^{n-1}\xone\sqrt{-2k_1^2}
\,\,\ell_1^{\mu_1}\ell_1^{\mu_2}\cdots\ell_1^{\mu_n}
\left(G_{a_1a_1\ldots a_n}(\xi_1,\xi_2,\ldots,\xi_n)\right)_{ij}
~,
\end{equation}
%
where we abreviate
%
\begin{equation}
\xi_i=\lid{\ell_1}{p_i}
~,
\end{equation}
%
and where
%
\begin{equation}
{G}_{a_1a_2\cdots a_n}(\xi_1,\xi_2,\ldots,\xi_n)
=
(-1)^{n-1}\sum_{\mathrm{perm.}}
\frac{T^{a_1}T^{a_2}\cdots T^{a_n}}
     {\xi_1(\xi_1+\xi_2)\cdots(\xi_1+\xi_2+\cdots+\xi_{n-1})}
~,
\end{equation}
%
which is identical to equation~(19) in~\cite{Antonov:2004hh}.
The sum is over all permutations of $(1,2,\ldots,n)$.
Due to momentum conservation and $\lid{\ell_1}{k_1}=0$, the arguments to ${G}_{a_1a_2\cdots a_n}$ are restricted such that $\xi_1+\xi_2+\cdots+\xi_{n-1}+\xi_n=0$.
In \Appendix{App:proof} we prove that this implies the identity
%
\begin{equation}
\Tr\left\{G_{a_1a_1\ldots a_n}(\xi_1,\xi_2,\ldots,\xi_n)\right\}=0
\label{Eqn:traceident}
~,
\end{equation}
%
so that
%
\begin{equation}
\left(G_{a_1a_1\ldots a_n}(\xi_1,\xi_2,\ldots,\xi_n)\right)_{ij}
=
2T^c_{ij}\,\Tr\left\{T^c\,G_{a_1a_1\ldots a_n}(\xi_1,\xi_2,\ldots,\xi_n)\right\}
~.
\end{equation}
%
Combining the foregoing, we see that
%
\begin{equation}
\raisebox{-28pt}{\epsfig{figure=graph42.eps,width=0.2\linewidth}}
=
\frac{-\xone E}{\sqrt{-k_1^2}}\,\sqrt{2}T^c_{ij}\left(-\imag\gqcd^{n-1}\right)
\Delta_{ca_1a_2\cdots a_n}^{-\mu_1\mu_2\cdots\mu_n}(\xi_1,\xi_2,\ldots,\xi_n)
~,
\label{eqn:effective}
\end{equation}
%
with
%
\begin{equation}
\Delta_{ca_1a_2\cdots a_n}^{-\mu_1\mu_2\cdots\mu_n}(\xi_1,\xi_2,\ldots,\xi_n)
=
-\frac{2k_1^2}{E}\,\ell_1^{\mu_1}\ell_1^{\mu_2}\cdots\ell_1^{\mu_n}\,
\Tr\left\{T^c\,G_{a_1a_1\ldots a_n}(\xi_1,\xi_2,\ldots,\xi_n)\right\}
~,
\end{equation}
%
which is the equivalent of equation~(18) in~\cite{Antonov:2004hh} for the eikonal $(-)$-direction.
The factor $E$ in the denominator is necessary to compensate for the fact that the vector $\ell_1$ has the dimension of $E$, while the vector $n^-$ in~\cite{Antonov:2004hh} is dimensionless.
For $n=3$, the factor $\imag\gqcd^2$ in \Equation{eqn:effective} corresponds to the same factor in the denominator on the l.h.s.\ of equation~(43) in~\cite{Antonov:2004hh}.
The remaining minus sign is due to the fact that in the formula above, all momenta are considered to be incoming, while $p_1,p_2$ in equation~(43) in~\cite{Antonov:2004hh} are outgoing.
Equation~(41) in~\cite{Antonov:2004hh} contains a misprint, and the factor $-\imag$ should be accompanying the $\Delta$s.

The vertices of \Figure{Fig4} contain all eikonal vertices and propagators.
Any vertex involving reggeized gluons can be constructed using these in combination with normal Feynman rules.
By recursively following the graphical decompositions of \Figure{Fig5} to \Figure{Fig8}, it can be understood that our prescription delivers the reggeon-reggeon-particles vertices up to $3$ particles.
On one hand, by expanding all blobs and effective vertices, one can see that all Feynman graphs for the process $q_A\,q_B\to q_A\,q_B\,X$ are obtained.
On the other hand, one can easily identify the graphs in our figures with those in figures 4, 5 and 8 from~\cite{Antonov:2004hh}.
Only for our \Figure{Fig8}, our notation is slightly more compact, and the comparison requires the expansion of the first term on the second line following \Figure{Fig7}.

\myFigure{%
\epsfig{figure=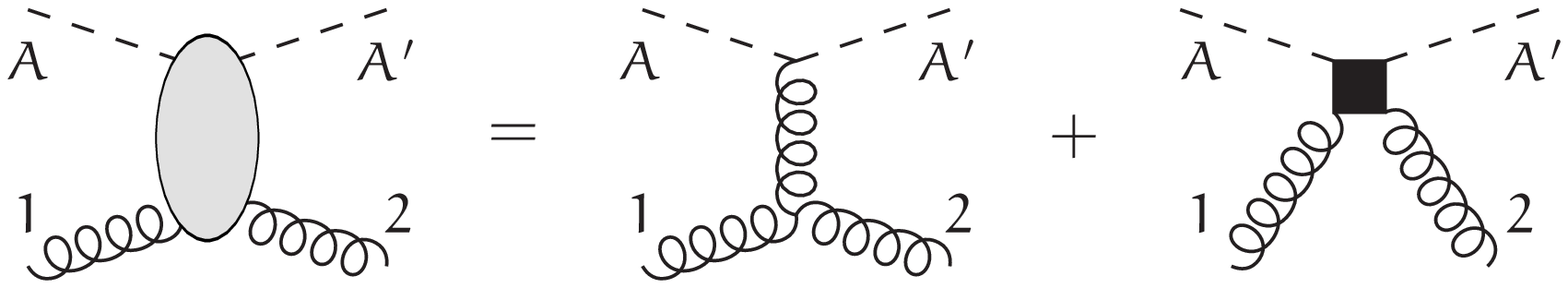,width=0.65\linewidth}
\caption{\label{Fig5}Graphs for the reggeon-gluon-gluon vertex.}
}
\myFigure{%
\epsfig{figure=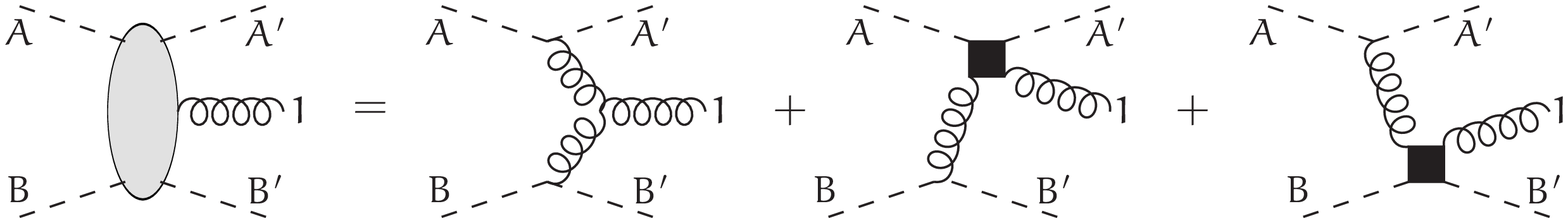,width=0.95\linewidth}\\[2ex]
\caption{\label{Fig6}Graphs for the reggeon-reggeon-gluon vertex.}
}
\myFigure{%
\epsfig{figure=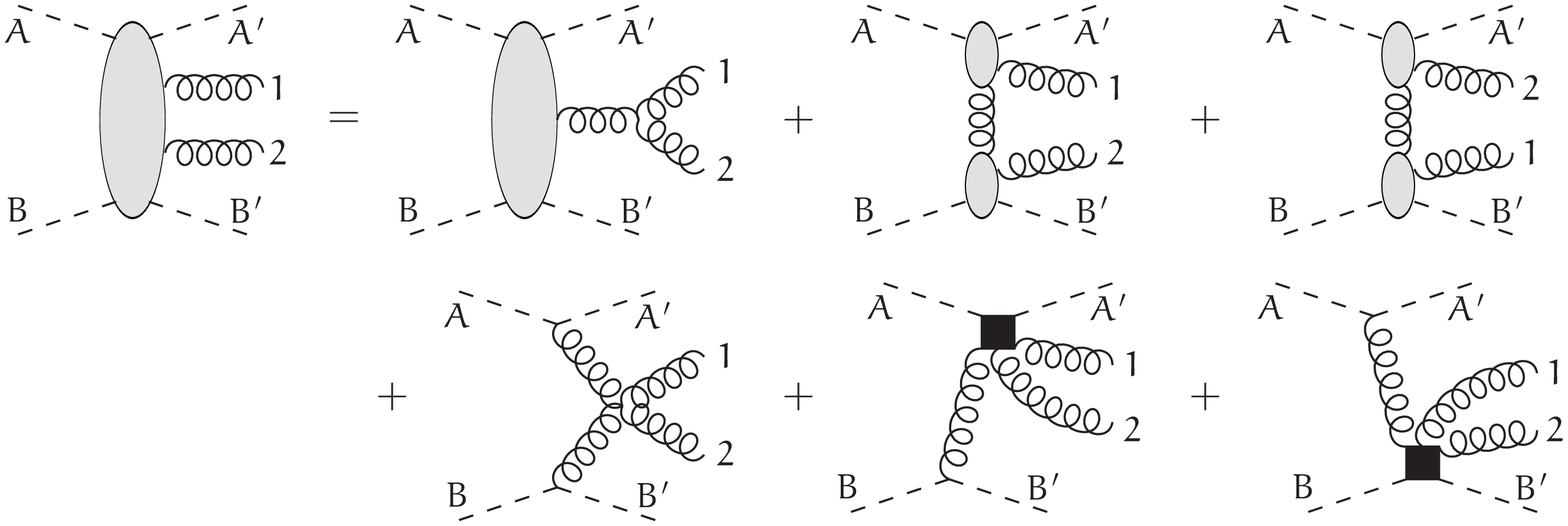,width=\linewidth}\\[2ex]
\caption{\label{Fig7}Graphs for the reggeon-reggeon-gluon-gluon vertex.}
}
\myFigure{%
\epsfig{figure=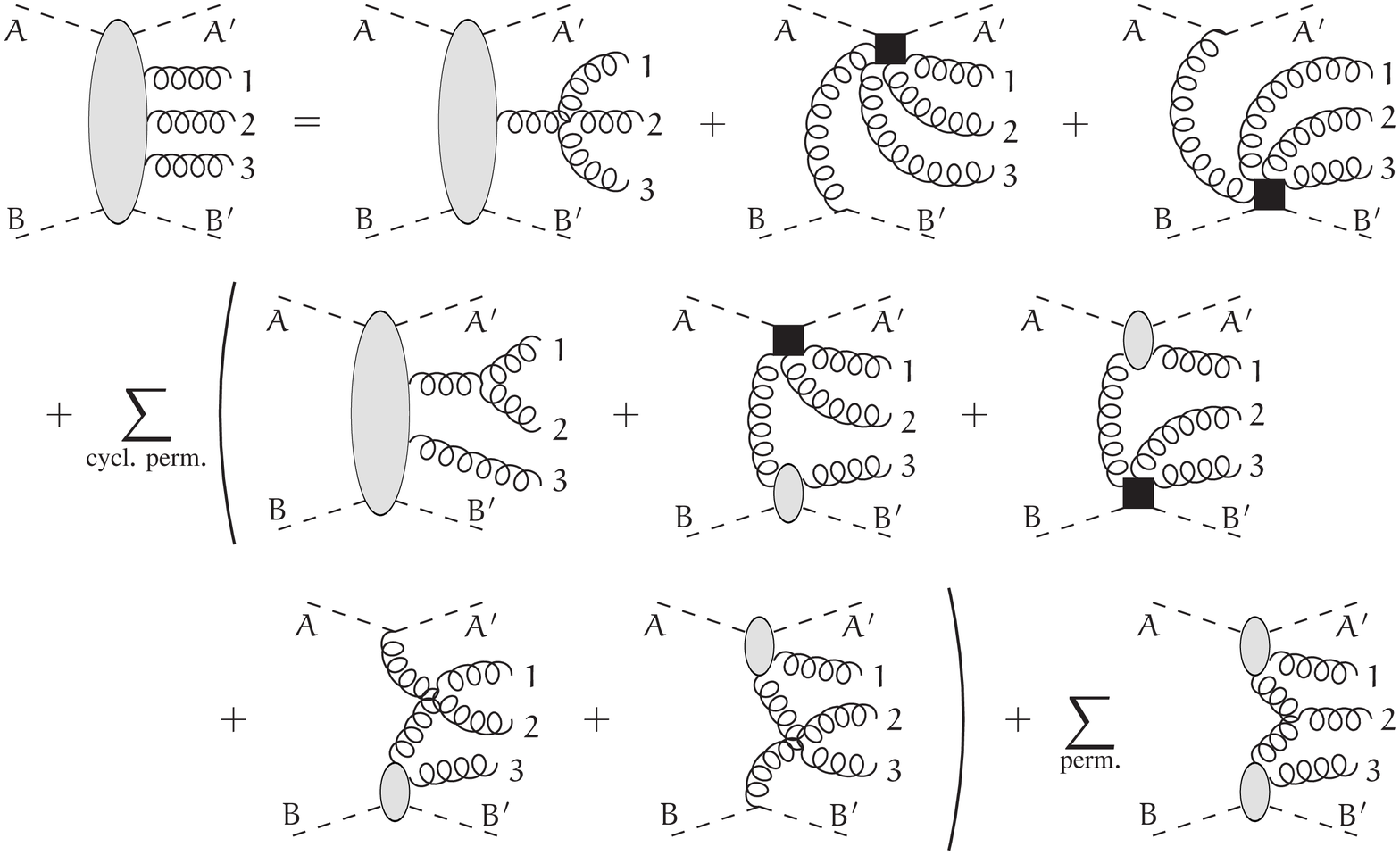,width=\linewidth}
\caption{\label{Fig8}Graphs for the reggeon-reggeon-gluon-gluon-gluon vertex.}
}

\section{\label{Sec:results}Numerical application}

We implemented the prescription of \Section{Sec:prescription} into a numerical program to calculate differential cross sections within high-energy factorization for arbitrary final states:
%
\begin{multline}
d\sigma
=
\int_{0}^{1}dx_1\int d^2k_{1\perp}\,f(x_1,k_{1\perp})
\int_{0}^{1}dx_2\int d^2k_{2\perp}\,f(x_2,k_{2\perp})
\,d\Phi(k_1,k_2;p_1,p_2,\ldots,p_n)
\\
\times
\frac{1}{(N_c^2-1)^2}\sum_{\mathrm{colors}}\sum_{\mathrm{helicities}}
\frac{\left|\Amp\big(g^*(k_1)\,g^*(k_2)\to X(p_1,p_2,\ldots,p_n)\big)\right|^2}
     {2s\,x_1^2x_2^2\,\mathrm{Sym}(X)}
~,
\end{multline}
%
where $k_1,k_2$ are defined as in \Equation{eqn:k1k2} and the sums are over all final-state helicities, and all colors, including the initial-state colors.
$\mathrm{Sym}(X)$ is the symmetry factor associated with possible identical particles in the final state, and $s$ is the square of the total center-of-mass energy.
The function~$f$ is the transversal-momentum dependent gluon density function.

Tree-level helicity amplitudes $\Amp$ are automatically calculated following the recursive Dyson-Schwinger method, in the spirit of~\cite{Mangano:2002ea,Kanaki:2000ey,Moretti:2001zz,Gleisberg:2008fv,Kleiss:2010hy}.
The helicity and color sums are performed numerically, evaluating so-called color-dressed helicity amplitudes~\cite{Papadopoulos:2005ky,Duhr:2006iq}.
As mentioned at the beginning, this happens following the prescription of \Section{Sec:prescription}, \ie, the recursive Dyson-Schwinger relations are augmented with the eikonal Feynman rules, and amplitudes with two initial-state and two final-state eikonal quarks are evaluated.
This increases the multiplicity by $2$, but this causes no serious burden for tree-level processes up to $6$ final-state particles.
Also, the eikonal vertices and propagators are computationally less involved than the other vertices and propagators.
The phase space integration of the final-state momenta is performed with {\sc Kaleu}~\cite{vanHameren:2010gg}, and the integration over the initial-state variables $x_1,x_2,k_{1\perp},k_{2\perp}$ is performed with adaptive grids following the method of~\cite{vanHameren:2007pt}.

Since the scope of this paper is rather a proof of concept than a phenomenological study, we choose to present results within a simple model for the transversal-momentum dependent pdf.
We use
\begin{equation}
f(x,k_\perp) = \theta(\mu-k_\perp)\,\frac{1}{2\pi\,Q_0^2}\,\exp\left(-\frac{k_\perp^2}{2Q_0^2\,x\,f_\mathrm{col}(x,\mu)}\right)
~,
\label{eqn:defUnint}
\end{equation}
%
where $Q_0$ is a scale of a few GeV, and $f_\mathrm{col}(x,\mu)$ is the collinear gluon density from CTEQ6L1~\cite{Pumplin:2002vw}.
For simplicity, we just fix the scale $\mu$ to the mass of the $Z$-boson.
\Equation{eqn:defUnint} has the property that the off-shellness of the initial-state gluon is squeezed to zero for small values of $Q_0$, such that
%
\begin{equation}
\lim_{Q_0\to0}\int d^2k_{\perp}\,f(x,k_\perp) = x\,f_\mathrm{col}(x,\mu)
~.
\end{equation}
%
This way, we can check the collinear limit without having to worry about the collinear limit of the transversal-momentum dependent pdf.

In the following, we present some differential distributions for the processes
\begin{align}
g^*g^* &\to b\bar{b}Z\to b\bar{b}\,\mu^+\mu^- \\
g^*g^* &\to b\bar{b}Zg\to b\bar{b}\,\mu^+\mu^-\,g \\
g^*g^* &\to b\bar{b}g \\
g^*g^* &\to b\bar{b}gg
~.
\end{align}
%
The Standard Model input parameters used are
%
\begin{gather}
m_W=80.419\,\GeV \quad \Gamma_W=2.12\,\GeV \\
m_Z=91.1882\,\GeV \quad \Gamma_Z=2.4952\,\GeV \\
G_\mu=1.16639\times10^{-5}\,\GeV^{-2} \\
\alpha_S=0.13
\end{gather}
The value of the fine-structure constant is defined following the $G_\mu$ scheme
\begin{gather}
\sin^2\theta_W=1-m_W^2/m_Z^2 \\
\alpha = \frac{\sqrt{2}\,G_\mu\,m_W^2\,\sin^2\theta_W}{\pi}
\end{gather}
The masses of the fermions involved in the processes are
\begin{gather}
m_b=4.62\,\GeV \\
m_\mu=0.1056583\,\GeV
\end{gather}
%
All calculations are performed at a center-of-mass energy of $14\,\mathrm{TeV}$.
The phase space of gluons in the final state is restricted such that $p_T>1\,\GeV$, and such that
%
\begin{equation}
\sqrt{\Delta\phi^2+\Delta y^2}>0.4
~,
\end{equation}
%
where $\Delta\phi$ is the difference in azimuthal angle and $\Delta y$ the difference in rapidity with any of the other final-state gluons or quarks.

\myFigure{%
\begin{minipage}[l]{0.49\linewidth}
\begingroup
  \makeatletter
  \providecommand\color[2][]{%
    \GenericError{(gnuplot) \space\space\space\@spaces}{%
      Package color not loaded in conjunction with
      terminal option `colourtext'%
    }{See the gnuplot documentation for explanation.%
    }{Either use 'blacktext' in gnuplot or load the package
      color.sty in LaTeX.}%
    \renewcommand\color[2][]{}%
  }%
  \providecommand\includegraphics[2][]{%
    \GenericError{(gnuplot) \space\space\space\@spaces}{%
      Package graphicx or graphics not loaded%
    }{See the gnuplot documentation for explanation.%
    }{The gnuplot epslatex terminal needs graphicx.sty or graphics.sty.}%
    \renewcommand\includegraphics[2][]{}%
  }%
  \providecommand\rotatebox[2]{#2}%
  \@ifundefined{ifGPcolor}{%
    \newif\ifGPcolor
    \GPcolortrue
  }{}%
  \@ifundefined{ifGPblacktext}{%
    \newif\ifGPblacktext
    \GPblacktexttrue
  }{}%
  \let\gplgaddtomacro\g@addto@macro
  \gdef\gplbacktext{}%
  \gdef\gplfronttext{}%
  \makeatother
  \ifGPblacktext
    \def\colorrgb#1{}%
    \def\colorgray#1{}%
  \else
    \ifGPcolor
      \def\colorrgb#1{\color[rgb]{#1}}%
      \def\colorgray#1{\color[gray]{#1}}%
      \expandafter\def\csname LTw\endcsname{\color{white}}%
      \expandafter\def\csname LTb\endcsname{\color{black}}%
      \expandafter\def\csname LTa\endcsname{\color{black}}%
      \expandafter\def\csname LT0\endcsname{\color[rgb]{1,0,0}}%
      \expandafter\def\csname LT1\endcsname{\color[rgb]{0,1,0}}%
      \expandafter\def\csname LT2\endcsname{\color[rgb]{0,0,1}}%
      \expandafter\def\csname LT3\endcsname{\color[rgb]{1,0,1}}%
      \expandafter\def\csname LT4\endcsname{\color[rgb]{0,1,1}}%
      \expandafter\def\csname LT5\endcsname{\color[rgb]{1,1,0}}%
      \expandafter\def\csname LT6\endcsname{\color[rgb]{0,0,0}}%
      \expandafter\def\csname LT7\endcsname{\color[rgb]{1,0.3,0}}%
      \expandafter\def\csname LT8\endcsname{\color[rgb]{0.5,0.5,0.5}}%
    \else
      \def\colorrgb#1{\color{black}}%
      \def\colorgray#1{\color[gray]{#1}}%
      \expandafter\def\csname LTw\endcsname{\color{white}}%
      \expandafter\def\csname LTb\endcsname{\color{black}}%
      \expandafter\def\csname LTa\endcsname{\color{black}}%
      \expandafter\def\csname LT0\endcsname{\color{black}}%
      \expandafter\def\csname LT1\endcsname{\color{black}}%
      \expandafter\def\csname LT2\endcsname{\color{black}}%
      \expandafter\def\csname LT3\endcsname{\color{black}}%
      \expandafter\def\csname LT4\endcsname{\color{black}}%
      \expandafter\def\csname LT5\endcsname{\color{black}}%
      \expandafter\def\csname LT6\endcsname{\color{black}}%
      \expandafter\def\csname LT7\endcsname{\color{black}}%
      \expandafter\def\csname LT8\endcsname{\color{black}}%
    \fi
  \fi
  \setlength{\unitlength}{0.0500bp}%
  \begin{picture}(4680.00,3528.00)%
    \gplgaddtomacro\gplbacktext{%
      \csname LTb\endcsname%
      \put(946,704){\makebox(0,0)[r]{\strut{}$10^{-5}$}}%
      \put(946,1216){\makebox(0,0)[r]{\strut{}$10^{-4}$}}%
      \put(946,1728){\makebox(0,0)[r]{\strut{}$10^{-3}$}}%
      \put(946,2239){\makebox(0,0)[r]{\strut{}$10^{-2}$}}%
      \put(946,2751){\makebox(0,0)[r]{\strut{}$10^{-1}$}}%
      \put(946,3263){\makebox(0,0)[r]{\strut{}$10^{0}$}}%
      \put(1078,484){\makebox(0,0){\strut{} 0}}%
      \put(1879,484){\makebox(0,0){\strut{} 100}}%
      \put(2681,484){\makebox(0,0){\strut{} 200}}%
      \put(3482,484){\makebox(0,0){\strut{} 300}}%
      \put(4283,484){\makebox(0,0){\strut{} 400}}%
      \put(176,1983){\rotatebox{-270}{\makebox(0,0){\strut{}$d\sigma/dp_T(Z)$ [pb/GeV]}}}%
      \put(2680,154){\makebox(0,0){\strut{}$p_T(Z)$ [GeV]}}%
      \put(1238,1062){\makebox(0,0)[l]{\strut{}$g^*g^*\to b\bar{b}Z\to b\bar{b}\mu^+\mu^-$}}%
    }%
    \gplgaddtomacro\gplfronttext{%
      \csname LTb\endcsname%
      \put(3296,3057){\makebox(0,0)[r]{\strut{}$Q_0=4$ GeV}}%
      \csname LTb\endcsname%
      \put(3296,2771){\makebox(0,0)[r]{\strut{}$Q_0=0$ GeV}}%
    }%
    \gplbacktext
    \put(0,0){\includegraphics{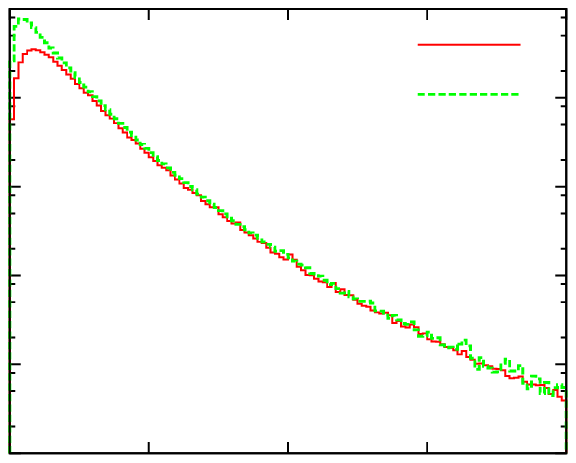}}%
    \gplfronttext
  \end{picture}%
\endgroup

\end{minipage}
\begin{minipage}[r]{0.49\linewidth}
\begingroup
  \makeatletter
  \providecommand\color[2][]{%
    \GenericError{(gnuplot) \space\space\space\@spaces}{%
      Package color not loaded in conjunction with
      terminal option `colourtext'%
    }{See the gnuplot documentation for explanation.%
    }{Either use 'blacktext' in gnuplot or load the package
      color.sty in LaTeX.}%
    \renewcommand\color[2][]{}%
  }%
  \providecommand\includegraphics[2][]{%
    \GenericError{(gnuplot) \space\space\space\@spaces}{%
      Package graphicx or graphics not loaded%
    }{See the gnuplot documentation for explanation.%
    }{The gnuplot epslatex terminal needs graphicx.sty or graphics.sty.}%
    \renewcommand\includegraphics[2][]{}%
  }%
  \providecommand\rotatebox[2]{#2}%
  \@ifundefined{ifGPcolor}{%
    \newif\ifGPcolor
    \GPcolortrue
  }{}%
  \@ifundefined{ifGPblacktext}{%
    \newif\ifGPblacktext
    \GPblacktexttrue
  }{}%
  \let\gplgaddtomacro\g@addto@macro
  \gdef\gplbacktext{}%
  \gdef\gplfronttext{}%
  \makeatother
  \ifGPblacktext
    \def\colorrgb#1{}%
    \def\colorgray#1{}%
  \else
    \ifGPcolor
      \def\colorrgb#1{\color[rgb]{#1}}%
      \def\colorgray#1{\color[gray]{#1}}%
      \expandafter\def\csname LTw\endcsname{\color{white}}%
      \expandafter\def\csname LTb\endcsname{\color{black}}%
      \expandafter\def\csname LTa\endcsname{\color{black}}%
      \expandafter\def\csname LT0\endcsname{\color[rgb]{1,0,0}}%
      \expandafter\def\csname LT1\endcsname{\color[rgb]{0,1,0}}%
      \expandafter\def\csname LT2\endcsname{\color[rgb]{0,0,1}}%
      \expandafter\def\csname LT3\endcsname{\color[rgb]{1,0,1}}%
      \expandafter\def\csname LT4\endcsname{\color[rgb]{0,1,1}}%
      \expandafter\def\csname LT5\endcsname{\color[rgb]{1,1,0}}%
      \expandafter\def\csname LT6\endcsname{\color[rgb]{0,0,0}}%
      \expandafter\def\csname LT7\endcsname{\color[rgb]{1,0.3,0}}%
      \expandafter\def\csname LT8\endcsname{\color[rgb]{0.5,0.5,0.5}}%
    \else
      \def\colorrgb#1{\color{black}}%
      \def\colorgray#1{\color[gray]{#1}}%
      \expandafter\def\csname LTw\endcsname{\color{white}}%
      \expandafter\def\csname LTb\endcsname{\color{black}}%
      \expandafter\def\csname LTa\endcsname{\color{black}}%
      \expandafter\def\csname LT0\endcsname{\color{black}}%
      \expandafter\def\csname LT1\endcsname{\color{black}}%
      \expandafter\def\csname LT2\endcsname{\color{black}}%
      \expandafter\def\csname LT3\endcsname{\color{black}}%
      \expandafter\def\csname LT4\endcsname{\color{black}}%
      \expandafter\def\csname LT5\endcsname{\color{black}}%
      \expandafter\def\csname LT6\endcsname{\color{black}}%
      \expandafter\def\csname LT7\endcsname{\color{black}}%
      \expandafter\def\csname LT8\endcsname{\color{black}}%
    \fi
  \fi
  \setlength{\unitlength}{0.0500bp}%
  \begin{picture}(4680.00,3528.00)%
    \gplgaddtomacro\gplbacktext{%
      \csname LTb\endcsname%
      \put(682,704){\makebox(0,0)[r]{\strut{} 0}}%
      \put(682,1070){\makebox(0,0)[r]{\strut{} 1}}%
      \put(682,1435){\makebox(0,0)[r]{\strut{} 2}}%
      \put(682,1801){\makebox(0,0)[r]{\strut{} 3}}%
      \put(682,2166){\makebox(0,0)[r]{\strut{} 4}}%
      \put(682,2532){\makebox(0,0)[r]{\strut{} 5}}%
      \put(682,2897){\makebox(0,0)[r]{\strut{} 6}}%
      \put(682,3263){\makebox(0,0)[r]{\strut{} 7}}%
      \put(814,484){\makebox(0,0){\strut{}-6}}%
      \put(1392,484){\makebox(0,0){\strut{}-4}}%
      \put(1970,484){\makebox(0,0){\strut{}-2}}%
      \put(2549,484){\makebox(0,0){\strut{} 0}}%
      \put(3127,484){\makebox(0,0){\strut{} 2}}%
      \put(3705,484){\makebox(0,0){\strut{} 4}}%
      \put(4283,484){\makebox(0,0){\strut{} 6}}%
      \put(176,1983){\rotatebox{-270}{\makebox(0,0){\strut{}$d\sigma/dy(Z)$ [pb]}}}%
      \put(2548,154){\makebox(0,0){\strut{}$y(Z)$}}%
    }%
    \gplgaddtomacro\gplfronttext{%
    }%
    \gplbacktext
    \put(0,0){\includegraphics{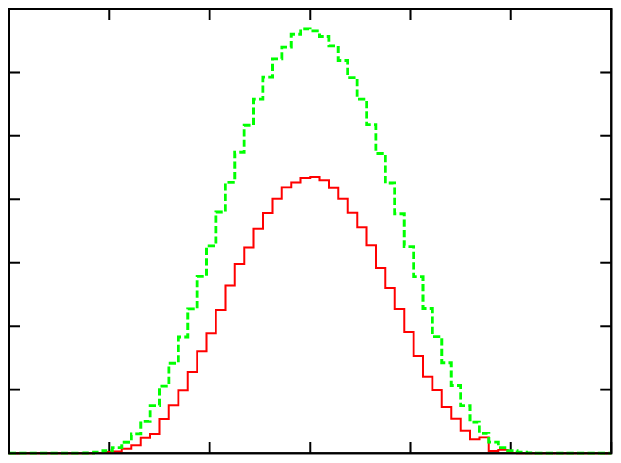}}%
    \gplfronttext
  \end{picture}%
\endgroup

\end{minipage}
\begin{minipage}[l]{0.49\linewidth}
\begingroup
  \makeatletter
  \providecommand\color[2][]{%
    \GenericError{(gnuplot) \space\space\space\@spaces}{%
      Package color not loaded in conjunction with
      terminal option `colourtext'%
    }{See the gnuplot documentation for explanation.%
    }{Either use 'blacktext' in gnuplot or load the package
      color.sty in LaTeX.}%
    \renewcommand\color[2][]{}%
  }%
  \providecommand\includegraphics[2][]{%
    \GenericError{(gnuplot) \space\space\space\@spaces}{%
      Package graphicx or graphics not loaded%
    }{See the gnuplot documentation for explanation.%
    }{The gnuplot epslatex terminal needs graphicx.sty or graphics.sty.}%
    \renewcommand\includegraphics[2][]{}%
  }%
  \providecommand\rotatebox[2]{#2}%
  \@ifundefined{ifGPcolor}{%
    \newif\ifGPcolor
    \GPcolortrue
  }{}%
  \@ifundefined{ifGPblacktext}{%
    \newif\ifGPblacktext
    \GPblacktexttrue
  }{}%
  \let\gplgaddtomacro\g@addto@macro
  \gdef\gplbacktext{}%
  \gdef\gplfronttext{}%
  \makeatother
  \ifGPblacktext
    \def\colorrgb#1{}%
    \def\colorgray#1{}%
  \else
    \ifGPcolor
      \def\colorrgb#1{\color[rgb]{#1}}%
      \def\colorgray#1{\color[gray]{#1}}%
      \expandafter\def\csname LTw\endcsname{\color{white}}%
      \expandafter\def\csname LTb\endcsname{\color{black}}%
      \expandafter\def\csname LTa\endcsname{\color{black}}%
      \expandafter\def\csname LT0\endcsname{\color[rgb]{1,0,0}}%
      \expandafter\def\csname LT1\endcsname{\color[rgb]{0,1,0}}%
      \expandafter\def\csname LT2\endcsname{\color[rgb]{0,0,1}}%
      \expandafter\def\csname LT3\endcsname{\color[rgb]{1,0,1}}%
      \expandafter\def\csname LT4\endcsname{\color[rgb]{0,1,1}}%
      \expandafter\def\csname LT5\endcsname{\color[rgb]{1,1,0}}%
      \expandafter\def\csname LT6\endcsname{\color[rgb]{0,0,0}}%
      \expandafter\def\csname LT7\endcsname{\color[rgb]{1,0.3,0}}%
      \expandafter\def\csname LT8\endcsname{\color[rgb]{0.5,0.5,0.5}}%
    \else
      \def\colorrgb#1{\color{black}}%
      \def\colorgray#1{\color[gray]{#1}}%
      \expandafter\def\csname LTw\endcsname{\color{white}}%
      \expandafter\def\csname LTb\endcsname{\color{black}}%
      \expandafter\def\csname LTa\endcsname{\color{black}}%
      \expandafter\def\csname LT0\endcsname{\color{black}}%
      \expandafter\def\csname LT1\endcsname{\color{black}}%
      \expandafter\def\csname LT2\endcsname{\color{black}}%
      \expandafter\def\csname LT3\endcsname{\color{black}}%
      \expandafter\def\csname LT4\endcsname{\color{black}}%
      \expandafter\def\csname LT5\endcsname{\color{black}}%
      \expandafter\def\csname LT6\endcsname{\color{black}}%
      \expandafter\def\csname LT7\endcsname{\color{black}}%
      \expandafter\def\csname LT8\endcsname{\color{black}}%
    \fi
  \fi
  \setlength{\unitlength}{0.0500bp}%
  \begin{picture}(4680.00,3528.00)%
    \gplgaddtomacro\gplbacktext{%
      \csname LTb\endcsname%
      \put(946,704){\makebox(0,0)[r]{\strut{}$10^{-5}$}}%
      \put(946,1216){\makebox(0,0)[r]{\strut{}$10^{-4}$}}%
      \put(946,1728){\makebox(0,0)[r]{\strut{}$10^{-3}$}}%
      \put(946,2239){\makebox(0,0)[r]{\strut{}$10^{-2}$}}%
      \put(946,2751){\makebox(0,0)[r]{\strut{}$10^{-1}$}}%
      \put(946,3263){\makebox(0,0)[r]{\strut{}$10^{0}$}}%
      \put(1078,484){\makebox(0,0){\strut{} 0}}%
      \put(1879,484){\makebox(0,0){\strut{} 100}}%
      \put(2681,484){\makebox(0,0){\strut{} 200}}%
      \put(3482,484){\makebox(0,0){\strut{} 300}}%
      \put(4283,484){\makebox(0,0){\strut{} 400}}%
      \put(176,1983){\rotatebox{-270}{\makebox(0,0){\strut{}$d\sigma/dp_T(b\bar{b})$ [pb/GeV]}}}%
      \put(2680,154){\makebox(0,0){\strut{}$p_T(b\bar{b})$ [GeV]}}%
      \put(1238,1062){\makebox(0,0)[l]{\strut{}$g^*g^*\to b\bar{b}Z\to b\bar{b}\mu^+\mu^-$}}%
    }%
    \gplgaddtomacro\gplfronttext{%
      \csname LTb\endcsname%
      \put(3296,3057){\makebox(0,0)[r]{\strut{}$Q_0=4$ GeV}}%
      \csname LTb\endcsname%
      \put(3296,2771){\makebox(0,0)[r]{\strut{}$Q_0=0$ GeV}}%
    }%
    \gplbacktext
    \put(0,0){\includegraphics{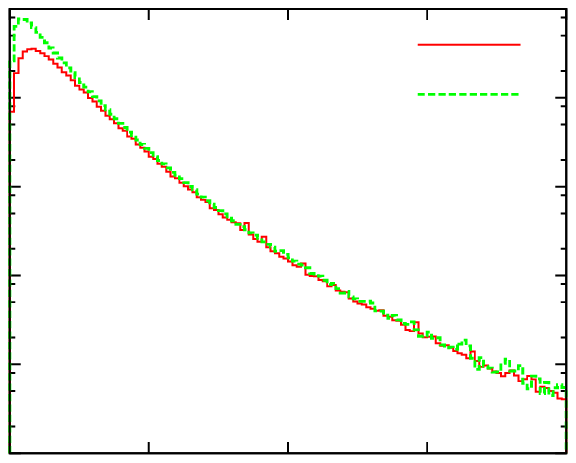}}%
    \gplfronttext
  \end{picture}%
\endgroup

\end{minipage}
\begin{minipage}[r]{0.49\linewidth}
\begingroup
  \makeatletter
  \providecommand\color[2][]{%
    \GenericError{(gnuplot) \space\space\space\@spaces}{%
      Package color not loaded in conjunction with
      terminal option `colourtext'%
    }{See the gnuplot documentation for explanation.%
    }{Either use 'blacktext' in gnuplot or load the package
      color.sty in LaTeX.}%
    \renewcommand\color[2][]{}%
  }%
  \providecommand\includegraphics[2][]{%
    \GenericError{(gnuplot) \space\space\space\@spaces}{%
      Package graphicx or graphics not loaded%
    }{See the gnuplot documentation for explanation.%
    }{The gnuplot epslatex terminal needs graphicx.sty or graphics.sty.}%
    \renewcommand\includegraphics[2][]{}%
  }%
  \providecommand\rotatebox[2]{#2}%
  \@ifundefined{ifGPcolor}{%
    \newif\ifGPcolor
    \GPcolortrue
  }{}%
  \@ifundefined{ifGPblacktext}{%
    \newif\ifGPblacktext
    \GPblacktexttrue
  }{}%
  \let\gplgaddtomacro\g@addto@macro
  \gdef\gplbacktext{}%
  \gdef\gplfronttext{}%
  \makeatother
  \ifGPblacktext
    \def\colorrgb#1{}%
    \def\colorgray#1{}%
  \else
    \ifGPcolor
      \def\colorrgb#1{\color[rgb]{#1}}%
      \def\colorgray#1{\color[gray]{#1}}%
      \expandafter\def\csname LTw\endcsname{\color{white}}%
      \expandafter\def\csname LTb\endcsname{\color{black}}%
      \expandafter\def\csname LTa\endcsname{\color{black}}%
      \expandafter\def\csname LT0\endcsname{\color[rgb]{1,0,0}}%
      \expandafter\def\csname LT1\endcsname{\color[rgb]{0,1,0}}%
      \expandafter\def\csname LT2\endcsname{\color[rgb]{0,0,1}}%
      \expandafter\def\csname LT3\endcsname{\color[rgb]{1,0,1}}%
      \expandafter\def\csname LT4\endcsname{\color[rgb]{0,1,1}}%
      \expandafter\def\csname LT5\endcsname{\color[rgb]{1,1,0}}%
      \expandafter\def\csname LT6\endcsname{\color[rgb]{0,0,0}}%
      \expandafter\def\csname LT7\endcsname{\color[rgb]{1,0.3,0}}%
      \expandafter\def\csname LT8\endcsname{\color[rgb]{0.5,0.5,0.5}}%
    \else
      \def\colorrgb#1{\color{black}}%
      \def\colorgray#1{\color[gray]{#1}}%
      \expandafter\def\csname LTw\endcsname{\color{white}}%
      \expandafter\def\csname LTb\endcsname{\color{black}}%
      \expandafter\def\csname LTa\endcsname{\color{black}}%
      \expandafter\def\csname LT0\endcsname{\color{black}}%
      \expandafter\def\csname LT1\endcsname{\color{black}}%
      \expandafter\def\csname LT2\endcsname{\color{black}}%
      \expandafter\def\csname LT3\endcsname{\color{black}}%
      \expandafter\def\csname LT4\endcsname{\color{black}}%
      \expandafter\def\csname LT5\endcsname{\color{black}}%
      \expandafter\def\csname LT6\endcsname{\color{black}}%
      \expandafter\def\csname LT7\endcsname{\color{black}}%
      \expandafter\def\csname LT8\endcsname{\color{black}}%
    \fi
  \fi
  \setlength{\unitlength}{0.0500bp}%
  \begin{picture}(4680.00,3528.00)%
    \gplgaddtomacro\gplbacktext{%
      \csname LTb\endcsname%
      \put(682,704){\makebox(0,0)[r]{\strut{} 0}}%
      \put(682,1070){\makebox(0,0)[r]{\strut{} 1}}%
      \put(682,1435){\makebox(0,0)[r]{\strut{} 2}}%
      \put(682,1801){\makebox(0,0)[r]{\strut{} 3}}%
      \put(682,2166){\makebox(0,0)[r]{\strut{} 4}}%
      \put(682,2532){\makebox(0,0)[r]{\strut{} 5}}%
      \put(682,2897){\makebox(0,0)[r]{\strut{} 6}}%
      \put(682,3263){\makebox(0,0)[r]{\strut{} 7}}%
      \put(814,484){\makebox(0,0){\strut{}-6}}%
      \put(1392,484){\makebox(0,0){\strut{}-4}}%
      \put(1970,484){\makebox(0,0){\strut{}-2}}%
      \put(2549,484){\makebox(0,0){\strut{} 0}}%
      \put(3127,484){\makebox(0,0){\strut{} 2}}%
      \put(3705,484){\makebox(0,0){\strut{} 4}}%
      \put(4283,484){\makebox(0,0){\strut{} 6}}%
      \put(176,1983){\rotatebox{-270}{\makebox(0,0){\strut{}$d\sigma/dy(b\bar{b})$ [pb]}}}%
      \put(2548,154){\makebox(0,0){\strut{}$y(b\bar{b})$}}%
    }%
    \gplgaddtomacro\gplfronttext{%
    }%
    \gplbacktext
    \put(0,0){\includegraphics{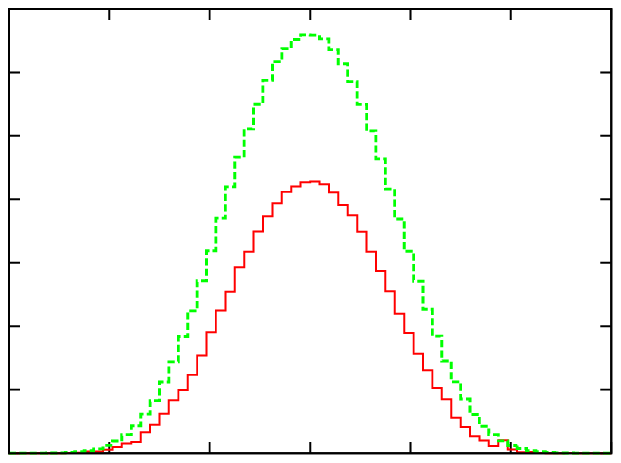}}%
    \gplfronttext
  \end{picture}%
\endgroup

\end{minipage}
\caption{\label{fig:bBlL}Differential cross sections of the transverse momentum $p_T$ and the rapidity $y$ of the $Z$-boson and the $b\bar{b}$-pair in the process $g^*g^*\to b\bar{b}Z\to b\bar{b}\mu^+\mu^-$. The different curves correspond to the different values of $Q_0$ in \Equation{eqn:defUnint}, with $Q_0=0$ refering to collinear factorization. The total cross sections are
$26.181\pm0.014\,\mathrm{pb}$ for $Q_0=0\,\GeV$,
$16.159\pm0.036\,\mathrm{pb}$ for $Q_0=4\,\GeV$.
These numbers are considerably smaller than those in \cite{Deak:2008ky}, because here the $Z$-boson is off-shell with only the $\mu^+\mu^-$-channel taken into account.}
}

\myFigure{%
\begin{minipage}[l]{0.49\linewidth}
\begingroup
  \makeatletter
  \providecommand\color[2][]{%
    \GenericError{(gnuplot) \space\space\space\@spaces}{%
      Package color not loaded in conjunction with
      terminal option `colourtext'%
    }{See the gnuplot documentation for explanation.%
    }{Either use 'blacktext' in gnuplot or load the package
      color.sty in LaTeX.}%
    \renewcommand\color[2][]{}%
  }%
  \providecommand\includegraphics[2][]{%
    \GenericError{(gnuplot) \space\space\space\@spaces}{%
      Package graphicx or graphics not loaded%
    }{See the gnuplot documentation for explanation.%
    }{The gnuplot epslatex terminal needs graphicx.sty or graphics.sty.}%
    \renewcommand\includegraphics[2][]{}%
  }%
  \providecommand\rotatebox[2]{#2}%
  \@ifundefined{ifGPcolor}{%
    \newif\ifGPcolor
    \GPcolortrue
  }{}%
  \@ifundefined{ifGPblacktext}{%
    \newif\ifGPblacktext
    \GPblacktexttrue
  }{}%
  \let\gplgaddtomacro\g@addto@macro
  \gdef\gplbacktext{}%
  \gdef\gplfronttext{}%
  \makeatother
  \ifGPblacktext
    \def\colorrgb#1{}%
    \def\colorgray#1{}%
  \else
    \ifGPcolor
      \def\colorrgb#1{\color[rgb]{#1}}%
      \def\colorgray#1{\color[gray]{#1}}%
      \expandafter\def\csname LTw\endcsname{\color{white}}%
      \expandafter\def\csname LTb\endcsname{\color{black}}%
      \expandafter\def\csname LTa\endcsname{\color{black}}%
      \expandafter\def\csname LT0\endcsname{\color[rgb]{1,0,0}}%
      \expandafter\def\csname LT1\endcsname{\color[rgb]{0,1,0}}%
      \expandafter\def\csname LT2\endcsname{\color[rgb]{0,0,1}}%
      \expandafter\def\csname LT3\endcsname{\color[rgb]{1,0,1}}%
      \expandafter\def\csname LT4\endcsname{\color[rgb]{0,1,1}}%
      \expandafter\def\csname LT5\endcsname{\color[rgb]{1,1,0}}%
      \expandafter\def\csname LT6\endcsname{\color[rgb]{0,0,0}}%
      \expandafter\def\csname LT7\endcsname{\color[rgb]{1,0.3,0}}%
      \expandafter\def\csname LT8\endcsname{\color[rgb]{0.5,0.5,0.5}}%
    \else
      \def\colorrgb#1{\color{black}}%
      \def\colorgray#1{\color[gray]{#1}}%
      \expandafter\def\csname LTw\endcsname{\color{white}}%
      \expandafter\def\csname LTb\endcsname{\color{black}}%
      \expandafter\def\csname LTa\endcsname{\color{black}}%
      \expandafter\def\csname LT0\endcsname{\color{black}}%
      \expandafter\def\csname LT1\endcsname{\color{black}}%
      \expandafter\def\csname LT2\endcsname{\color{black}}%
      \expandafter\def\csname LT3\endcsname{\color{black}}%
      \expandafter\def\csname LT4\endcsname{\color{black}}%
      \expandafter\def\csname LT5\endcsname{\color{black}}%
      \expandafter\def\csname LT6\endcsname{\color{black}}%
      \expandafter\def\csname LT7\endcsname{\color{black}}%
      \expandafter\def\csname LT8\endcsname{\color{black}}%
    \fi
  \fi
  \setlength{\unitlength}{0.0500bp}%
  \begin{picture}(4680.00,3528.00)%
    \gplgaddtomacro\gplbacktext{%
      \csname LTb\endcsname%
      \put(946,704){\makebox(0,0)[r]{\strut{}$10^{-4}$}}%
      \put(946,1216){\makebox(0,0)[r]{\strut{}$10^{-3}$}}%
      \put(946,1728){\makebox(0,0)[r]{\strut{}$10^{-2}$}}%
      \put(946,2239){\makebox(0,0)[r]{\strut{}$10^{-1}$}}%
      \put(946,2751){\makebox(0,0)[r]{\strut{}$10^{0}$}}%
      \put(946,3263){\makebox(0,0)[r]{\strut{}$10^{1}$}}%
      \put(1078,484){\makebox(0,0){\strut{} 0}}%
      \put(1879,484){\makebox(0,0){\strut{} 100}}%
      \put(2681,484){\makebox(0,0){\strut{} 200}}%
      \put(3482,484){\makebox(0,0){\strut{} 300}}%
      \put(4283,484){\makebox(0,0){\strut{} 400}}%
      \put(176,1983){\rotatebox{-270}{\makebox(0,0){\strut{}$d\sigma/dp_T(Z)$ [pb/GeV]}}}%
      \put(2680,154){\makebox(0,0){\strut{}$p_T(Z)$ [GeV]}}%
      \put(1174,948){\makebox(0,0)[l]{\strut{}$g^*g^*\to b\bar{b}Zg\to b\bar{b}\mu^+\mu^-g$}}%
    }%
    \gplgaddtomacro\gplfronttext{%
      \csname LTb\endcsname%
      \put(3296,3057){\makebox(0,0)[r]{\strut{}$Q_0=4$ GeV}}%
      \csname LTb\endcsname%
      \put(3296,2771){\makebox(0,0)[r]{\strut{}$Q_0=0$ GeV}}%
    }%
    \gplbacktext
    \put(0,0){\includegraphics{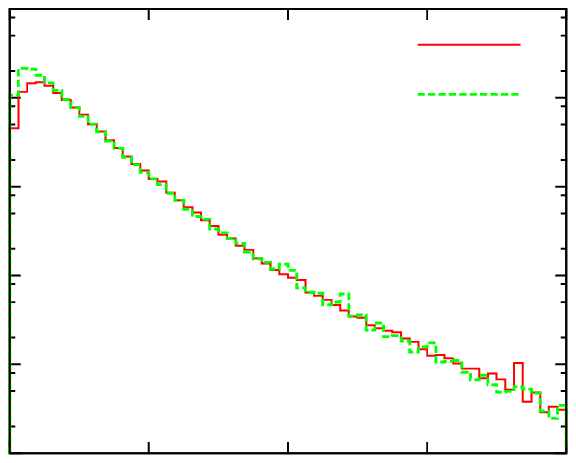}}%
    \gplfronttext
  \end{picture}%
\endgroup

\end{minipage}
\begin{minipage}[r]{0.49\linewidth}
\begingroup
  \makeatletter
  \providecommand\color[2][]{%
    \GenericError{(gnuplot) \space\space\space\@spaces}{%
      Package color not loaded in conjunction with
      terminal option `colourtext'%
    }{See the gnuplot documentation for explanation.%
    }{Either use 'blacktext' in gnuplot or load the package
      color.sty in LaTeX.}%
    \renewcommand\color[2][]{}%
  }%
  \providecommand\includegraphics[2][]{%
    \GenericError{(gnuplot) \space\space\space\@spaces}{%
      Package graphicx or graphics not loaded%
    }{See the gnuplot documentation for explanation.%
    }{The gnuplot epslatex terminal needs graphicx.sty or graphics.sty.}%
    \renewcommand\includegraphics[2][]{}%
  }%
  \providecommand\rotatebox[2]{#2}%
  \@ifundefined{ifGPcolor}{%
    \newif\ifGPcolor
    \GPcolortrue
  }{}%
  \@ifundefined{ifGPblacktext}{%
    \newif\ifGPblacktext
    \GPblacktexttrue
  }{}%
  \let\gplgaddtomacro\g@addto@macro
  \gdef\gplbacktext{}%
  \gdef\gplfronttext{}%
  \makeatother
  \ifGPblacktext
    \def\colorrgb#1{}%
    \def\colorgray#1{}%
  \else
    \ifGPcolor
      \def\colorrgb#1{\color[rgb]{#1}}%
      \def\colorgray#1{\color[gray]{#1}}%
      \expandafter\def\csname LTw\endcsname{\color{white}}%
      \expandafter\def\csname LTb\endcsname{\color{black}}%
      \expandafter\def\csname LTa\endcsname{\color{black}}%
      \expandafter\def\csname LT0\endcsname{\color[rgb]{1,0,0}}%
      \expandafter\def\csname LT1\endcsname{\color[rgb]{0,1,0}}%
      \expandafter\def\csname LT2\endcsname{\color[rgb]{0,0,1}}%
      \expandafter\def\csname LT3\endcsname{\color[rgb]{1,0,1}}%
      \expandafter\def\csname LT4\endcsname{\color[rgb]{0,1,1}}%
      \expandafter\def\csname LT5\endcsname{\color[rgb]{1,1,0}}%
      \expandafter\def\csname LT6\endcsname{\color[rgb]{0,0,0}}%
      \expandafter\def\csname LT7\endcsname{\color[rgb]{1,0.3,0}}%
      \expandafter\def\csname LT8\endcsname{\color[rgb]{0.5,0.5,0.5}}%
    \else
      \def\colorrgb#1{\color{black}}%
      \def\colorgray#1{\color[gray]{#1}}%
      \expandafter\def\csname LTw\endcsname{\color{white}}%
      \expandafter\def\csname LTb\endcsname{\color{black}}%
      \expandafter\def\csname LTa\endcsname{\color{black}}%
      \expandafter\def\csname LT0\endcsname{\color{black}}%
      \expandafter\def\csname LT1\endcsname{\color{black}}%
      \expandafter\def\csname LT2\endcsname{\color{black}}%
      \expandafter\def\csname LT3\endcsname{\color{black}}%
      \expandafter\def\csname LT4\endcsname{\color{black}}%
      \expandafter\def\csname LT5\endcsname{\color{black}}%
      \expandafter\def\csname LT6\endcsname{\color{black}}%
      \expandafter\def\csname LT7\endcsname{\color{black}}%
      \expandafter\def\csname LT8\endcsname{\color{black}}%
    \fi
  \fi
  \setlength{\unitlength}{0.0500bp}%
  \begin{picture}(4680.00,3528.00)%
    \gplgaddtomacro\gplbacktext{%
      \csname LTb\endcsname%
      \put(814,704){\makebox(0,0)[r]{\strut{} 0}}%
      \put(814,1216){\makebox(0,0)[r]{\strut{} 5}}%
      \put(814,1728){\makebox(0,0)[r]{\strut{} 10}}%
      \put(814,2239){\makebox(0,0)[r]{\strut{} 15}}%
      \put(814,2751){\makebox(0,0)[r]{\strut{} 20}}%
      \put(814,3263){\makebox(0,0)[r]{\strut{} 25}}%
      \put(946,484){\makebox(0,0){\strut{}-6}}%
      \put(1502,484){\makebox(0,0){\strut{}-4}}%
      \put(2058,484){\makebox(0,0){\strut{}-2}}%
      \put(2615,484){\makebox(0,0){\strut{} 0}}%
      \put(3171,484){\makebox(0,0){\strut{} 2}}%
      \put(3727,484){\makebox(0,0){\strut{} 4}}%
      \put(4283,484){\makebox(0,0){\strut{} 6}}%
      \put(176,1983){\rotatebox{-270}{\makebox(0,0){\strut{}$d\sigma/dy(Z)$ [pb]}}}%
      \put(2614,154){\makebox(0,0){\strut{}$y(Z)$}}%
    }%
    \gplgaddtomacro\gplfronttext{%
    }%
    \gplbacktext
    \put(0,0){\includegraphics{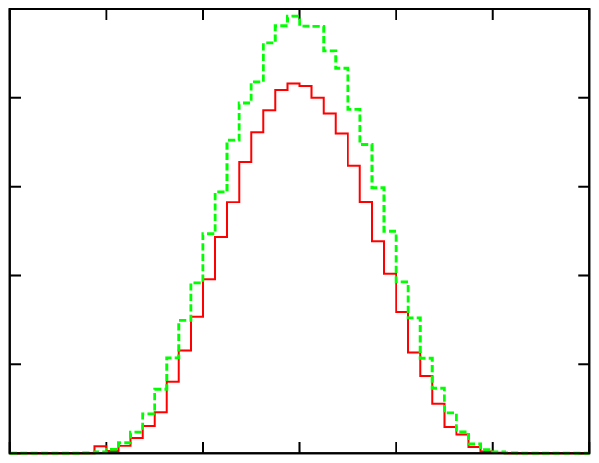}}%
    \gplfronttext
  \end{picture}%
\endgroup

\end{minipage}
\begin{minipage}[l]{0.49\linewidth}
\begingroup
  \makeatletter
  \providecommand\color[2][]{%
    \GenericError{(gnuplot) \space\space\space\@spaces}{%
      Package color not loaded in conjunction with
      terminal option `colourtext'%
    }{See the gnuplot documentation for explanation.%
    }{Either use 'blacktext' in gnuplot or load the package
      color.sty in LaTeX.}%
    \renewcommand\color[2][]{}%
  }%
  \providecommand\includegraphics[2][]{%
    \GenericError{(gnuplot) \space\space\space\@spaces}{%
      Package graphicx or graphics not loaded%
    }{See the gnuplot documentation for explanation.%
    }{The gnuplot epslatex terminal needs graphicx.sty or graphics.sty.}%
    \renewcommand\includegraphics[2][]{}%
  }%
  \providecommand\rotatebox[2]{#2}%
  \@ifundefined{ifGPcolor}{%
    \newif\ifGPcolor
    \GPcolortrue
  }{}%
  \@ifundefined{ifGPblacktext}{%
    \newif\ifGPblacktext
    \GPblacktexttrue
  }{}%
  \let\gplgaddtomacro\g@addto@macro
  \gdef\gplbacktext{}%
  \gdef\gplfronttext{}%
  \makeatother
  \ifGPblacktext
    \def\colorrgb#1{}%
    \def\colorgray#1{}%
  \else
    \ifGPcolor
      \def\colorrgb#1{\color[rgb]{#1}}%
      \def\colorgray#1{\color[gray]{#1}}%
      \expandafter\def\csname LTw\endcsname{\color{white}}%
      \expandafter\def\csname LTb\endcsname{\color{black}}%
      \expandafter\def\csname LTa\endcsname{\color{black}}%
      \expandafter\def\csname LT0\endcsname{\color[rgb]{1,0,0}}%
      \expandafter\def\csname LT1\endcsname{\color[rgb]{0,1,0}}%
      \expandafter\def\csname LT2\endcsname{\color[rgb]{0,0,1}}%
      \expandafter\def\csname LT3\endcsname{\color[rgb]{1,0,1}}%
      \expandafter\def\csname LT4\endcsname{\color[rgb]{0,1,1}}%
      \expandafter\def\csname LT5\endcsname{\color[rgb]{1,1,0}}%
      \expandafter\def\csname LT6\endcsname{\color[rgb]{0,0,0}}%
      \expandafter\def\csname LT7\endcsname{\color[rgb]{1,0.3,0}}%
      \expandafter\def\csname LT8\endcsname{\color[rgb]{0.5,0.5,0.5}}%
    \else
      \def\colorrgb#1{\color{black}}%
      \def\colorgray#1{\color[gray]{#1}}%
      \expandafter\def\csname LTw\endcsname{\color{white}}%
      \expandafter\def\csname LTb\endcsname{\color{black}}%
      \expandafter\def\csname LTa\endcsname{\color{black}}%
      \expandafter\def\csname LT0\endcsname{\color{black}}%
      \expandafter\def\csname LT1\endcsname{\color{black}}%
      \expandafter\def\csname LT2\endcsname{\color{black}}%
      \expandafter\def\csname LT3\endcsname{\color{black}}%
      \expandafter\def\csname LT4\endcsname{\color{black}}%
      \expandafter\def\csname LT5\endcsname{\color{black}}%
      \expandafter\def\csname LT6\endcsname{\color{black}}%
      \expandafter\def\csname LT7\endcsname{\color{black}}%
      \expandafter\def\csname LT8\endcsname{\color{black}}%
    \fi
  \fi
  \setlength{\unitlength}{0.0500bp}%
  \begin{picture}(4680.00,3528.00)%
    \gplgaddtomacro\gplbacktext{%
      \csname LTb\endcsname%
      \put(946,704){\makebox(0,0)[r]{\strut{}$10^{-4}$}}%
      \put(946,1216){\makebox(0,0)[r]{\strut{}$10^{-3}$}}%
      \put(946,1728){\makebox(0,0)[r]{\strut{}$10^{-2}$}}%
      \put(946,2239){\makebox(0,0)[r]{\strut{}$10^{-1}$}}%
      \put(946,2751){\makebox(0,0)[r]{\strut{}$10^{0}$}}%
      \put(946,3263){\makebox(0,0)[r]{\strut{}$10^{1}$}}%
      \put(1078,484){\makebox(0,0){\strut{} 0}}%
      \put(1879,484){\makebox(0,0){\strut{} 100}}%
      \put(2681,484){\makebox(0,0){\strut{} 200}}%
      \put(3482,484){\makebox(0,0){\strut{} 300}}%
      \put(4283,484){\makebox(0,0){\strut{} 400}}%
      \put(176,1983){\rotatebox{-270}{\makebox(0,0){\strut{}$d\sigma/dp_T(b\bar{b})$ [pb/GeV]}}}%
      \put(2680,154){\makebox(0,0){\strut{}$p_T(b\bar{b})$ [GeV]}}%
      \put(1174,948){\makebox(0,0)[l]{\strut{}$g^*g^*\to b\bar{b}Zg\to b\bar{b}\mu^+\mu^-g$}}%
    }%
    \gplgaddtomacro\gplfronttext{%
      \csname LTb\endcsname%
      \put(3296,3057){\makebox(0,0)[r]{\strut{}$Q_0=4$ GeV}}%
      \csname LTb\endcsname%
      \put(3296,2771){\makebox(0,0)[r]{\strut{}$Q_0=0$ GeV}}%
    }%
    \gplbacktext
    \put(0,0){\includegraphics{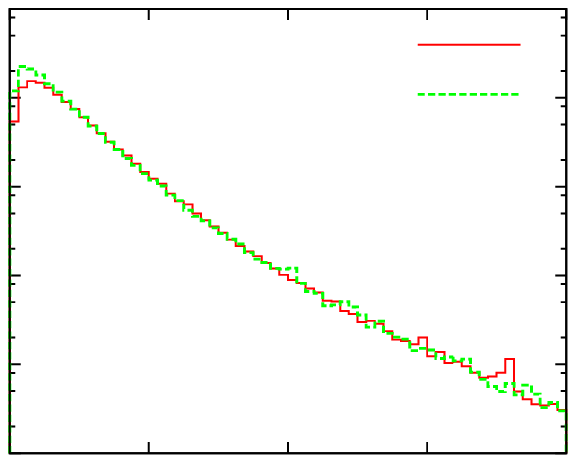}}%
    \gplfronttext
  \end{picture}%
\endgroup

\end{minipage}
\begin{minipage}[r]{0.49\linewidth}
\begingroup
  \makeatletter
  \providecommand\color[2][]{%
    \GenericError{(gnuplot) \space\space\space\@spaces}{%
      Package color not loaded in conjunction with
      terminal option `colourtext'%
    }{See the gnuplot documentation for explanation.%
    }{Either use 'blacktext' in gnuplot or load the package
      color.sty in LaTeX.}%
    \renewcommand\color[2][]{}%
  }%
  \providecommand\includegraphics[2][]{%
    \GenericError{(gnuplot) \space\space\space\@spaces}{%
      Package graphicx or graphics not loaded%
    }{See the gnuplot documentation for explanation.%
    }{The gnuplot epslatex terminal needs graphicx.sty or graphics.sty.}%
    \renewcommand\includegraphics[2][]{}%
  }%
  \providecommand\rotatebox[2]{#2}%
  \@ifundefined{ifGPcolor}{%
    \newif\ifGPcolor
    \GPcolortrue
  }{}%
  \@ifundefined{ifGPblacktext}{%
    \newif\ifGPblacktext
    \GPblacktexttrue
  }{}%
  \let\gplgaddtomacro\g@addto@macro
  \gdef\gplbacktext{}%
  \gdef\gplfronttext{}%
  \makeatother
  \ifGPblacktext
    \def\colorrgb#1{}%
    \def\colorgray#1{}%
  \else
    \ifGPcolor
      \def\colorrgb#1{\color[rgb]{#1}}%
      \def\colorgray#1{\color[gray]{#1}}%
      \expandafter\def\csname LTw\endcsname{\color{white}}%
      \expandafter\def\csname LTb\endcsname{\color{black}}%
      \expandafter\def\csname LTa\endcsname{\color{black}}%
      \expandafter\def\csname LT0\endcsname{\color[rgb]{1,0,0}}%
      \expandafter\def\csname LT1\endcsname{\color[rgb]{0,1,0}}%
      \expandafter\def\csname LT2\endcsname{\color[rgb]{0,0,1}}%
      \expandafter\def\csname LT3\endcsname{\color[rgb]{1,0,1}}%
      \expandafter\def\csname LT4\endcsname{\color[rgb]{0,1,1}}%
      \expandafter\def\csname LT5\endcsname{\color[rgb]{1,1,0}}%
      \expandafter\def\csname LT6\endcsname{\color[rgb]{0,0,0}}%
      \expandafter\def\csname LT7\endcsname{\color[rgb]{1,0.3,0}}%
      \expandafter\def\csname LT8\endcsname{\color[rgb]{0.5,0.5,0.5}}%
    \else
      \def\colorrgb#1{\color{black}}%
      \def\colorgray#1{\color[gray]{#1}}%
      \expandafter\def\csname LTw\endcsname{\color{white}}%
      \expandafter\def\csname LTb\endcsname{\color{black}}%
      \expandafter\def\csname LTa\endcsname{\color{black}}%
      \expandafter\def\csname LT0\endcsname{\color{black}}%
      \expandafter\def\csname LT1\endcsname{\color{black}}%
      \expandafter\def\csname LT2\endcsname{\color{black}}%
      \expandafter\def\csname LT3\endcsname{\color{black}}%
      \expandafter\def\csname LT4\endcsname{\color{black}}%
      \expandafter\def\csname LT5\endcsname{\color{black}}%
      \expandafter\def\csname LT6\endcsname{\color{black}}%
      \expandafter\def\csname LT7\endcsname{\color{black}}%
      \expandafter\def\csname LT8\endcsname{\color{black}}%
    \fi
  \fi
  \setlength{\unitlength}{0.0500bp}%
  \begin{picture}(4680.00,3528.00)%
    \gplgaddtomacro\gplbacktext{%
      \csname LTb\endcsname%
      \put(814,704){\makebox(0,0)[r]{\strut{} 0}}%
      \put(814,1216){\makebox(0,0)[r]{\strut{} 5}}%
      \put(814,1728){\makebox(0,0)[r]{\strut{} 10}}%
      \put(814,2239){\makebox(0,0)[r]{\strut{} 15}}%
      \put(814,2751){\makebox(0,0)[r]{\strut{} 20}}%
      \put(814,3263){\makebox(0,0)[r]{\strut{} 25}}%
      \put(946,484){\makebox(0,0){\strut{}-6}}%
      \put(1502,484){\makebox(0,0){\strut{}-4}}%
      \put(2058,484){\makebox(0,0){\strut{}-2}}%
      \put(2615,484){\makebox(0,0){\strut{} 0}}%
      \put(3171,484){\makebox(0,0){\strut{} 2}}%
      \put(3727,484){\makebox(0,0){\strut{} 4}}%
      \put(4283,484){\makebox(0,0){\strut{} 6}}%
      \put(176,1983){\rotatebox{-270}{\makebox(0,0){\strut{}$d\sigma/dy(b\bar{b})$ [pb]}}}%
      \put(2614,154){\makebox(0,0){\strut{}$y(b\bar{b})$}}%
    }%
    \gplgaddtomacro\gplfronttext{%
    }%
    \gplbacktext
    \put(0,0){\includegraphics{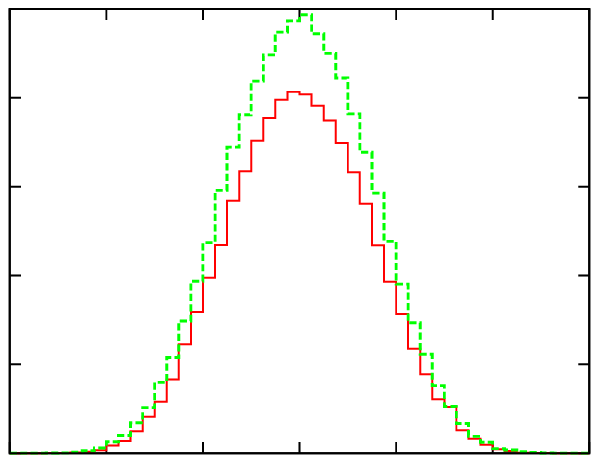}}%
    \gplfronttext
  \end{picture}%
\endgroup

\end{minipage}
\caption{\label{fig:bBZg}Differential cross sections of the transverse momentum $p_T$ and the rapidity $y$ of the $Z$-boson and the $b\bar{b}$-pair in the process $g^*g^*\to b\bar{b}Zg\to b\bar{b}\mu^+\mu^-g$. The different curves correspond to the different values of $Q_0$ in \Equation{eqn:defUnint}, with $Q_0=0$ refering to collinear factorization. The total cross sections are
$93.57\pm0.21\,\mathrm{pb}$ for $Q_0=0\,\GeV$,
$76.98\pm0.23\,\mathrm{pb}$ for $Q_0=4\,\GeV$.}
}

\myFigure{%
\begin{minipage}[l]{0.49\linewidth}
\begingroup
  \makeatletter
  \providecommand\color[2][]{%
    \GenericError{(gnuplot) \space\space\space\@spaces}{%
      Package color not loaded in conjunction with
      terminal option `colourtext'%
    }{See the gnuplot documentation for explanation.%
    }{Either use 'blacktext' in gnuplot or load the package
      color.sty in LaTeX.}%
    \renewcommand\color[2][]{}%
  }%
  \providecommand\includegraphics[2][]{%
    \GenericError{(gnuplot) \space\space\space\@spaces}{%
      Package graphicx or graphics not loaded%
    }{See the gnuplot documentation for explanation.%
    }{The gnuplot epslatex terminal needs graphicx.sty or graphics.sty.}%
    \renewcommand\includegraphics[2][]{}%
  }%
  \providecommand\rotatebox[2]{#2}%
  \@ifundefined{ifGPcolor}{%
    \newif\ifGPcolor
    \GPcolortrue
  }{}%
  \@ifundefined{ifGPblacktext}{%
    \newif\ifGPblacktext
    \GPblacktexttrue
  }{}%
  \let\gplgaddtomacro\g@addto@macro
  \gdef\gplbacktext{}%
  \gdef\gplfronttext{}%
  \makeatother
  \ifGPblacktext
    \def\colorrgb#1{}%
    \def\colorgray#1{}%
  \else
    \ifGPcolor
      \def\colorrgb#1{\color[rgb]{#1}}%
      \def\colorgray#1{\color[gray]{#1}}%
      \expandafter\def\csname LTw\endcsname{\color{white}}%
      \expandafter\def\csname LTb\endcsname{\color{black}}%
      \expandafter\def\csname LTa\endcsname{\color{black}}%
      \expandafter\def\csname LT0\endcsname{\color[rgb]{1,0,0}}%
      \expandafter\def\csname LT1\endcsname{\color[rgb]{0,1,0}}%
      \expandafter\def\csname LT2\endcsname{\color[rgb]{0,0,1}}%
      \expandafter\def\csname LT3\endcsname{\color[rgb]{1,0,1}}%
      \expandafter\def\csname LT4\endcsname{\color[rgb]{0,1,1}}%
      \expandafter\def\csname LT5\endcsname{\color[rgb]{1,1,0}}%
      \expandafter\def\csname LT6\endcsname{\color[rgb]{0,0,0}}%
      \expandafter\def\csname LT7\endcsname{\color[rgb]{1,0.3,0}}%
      \expandafter\def\csname LT8\endcsname{\color[rgb]{0.5,0.5,0.5}}%
    \else
      \def\colorrgb#1{\color{black}}%
      \def\colorgray#1{\color[gray]{#1}}%
      \expandafter\def\csname LTw\endcsname{\color{white}}%
      \expandafter\def\csname LTb\endcsname{\color{black}}%
      \expandafter\def\csname LTa\endcsname{\color{black}}%
      \expandafter\def\csname LT0\endcsname{\color{black}}%
      \expandafter\def\csname LT1\endcsname{\color{black}}%
      \expandafter\def\csname LT2\endcsname{\color{black}}%
      \expandafter\def\csname LT3\endcsname{\color{black}}%
      \expandafter\def\csname LT4\endcsname{\color{black}}%
      \expandafter\def\csname LT5\endcsname{\color{black}}%
      \expandafter\def\csname LT6\endcsname{\color{black}}%
      \expandafter\def\csname LT7\endcsname{\color{black}}%
      \expandafter\def\csname LT8\endcsname{\color{black}}%
    \fi
  \fi
  \setlength{\unitlength}{0.0500bp}%
  \begin{picture}(4680.00,3528.00)%
    \gplgaddtomacro\gplbacktext{%
      \csname LTb\endcsname%
      \put(946,704){\makebox(0,0)[r]{\strut{}$10^{-6}$}}%
      \put(946,988){\makebox(0,0)[r]{\strut{}$10^{-5}$}}%
      \put(946,1273){\makebox(0,0)[r]{\strut{}$10^{-4}$}}%
      \put(946,1557){\makebox(0,0)[r]{\strut{}$10^{-3}$}}%
      \put(946,1841){\makebox(0,0)[r]{\strut{}$10^{-2}$}}%
      \put(946,2126){\makebox(0,0)[r]{\strut{}$10^{-1}$}}%
      \put(946,2410){\makebox(0,0)[r]{\strut{}$10^{0}$}}%
      \put(946,2694){\makebox(0,0)[r]{\strut{}$10^{1}$}}%
      \put(946,2979){\makebox(0,0)[r]{\strut{}$10^{2}$}}%
      \put(946,3263){\makebox(0,0)[r]{\strut{}$10^{3}$}}%
      \put(1078,484){\makebox(0,0){\strut{} 0}}%
      \put(1879,484){\makebox(0,0){\strut{} 100}}%
      \put(2681,484){\makebox(0,0){\strut{} 200}}%
      \put(3482,484){\makebox(0,0){\strut{} 300}}%
      \put(4283,484){\makebox(0,0){\strut{} 400}}%
      \put(176,1983){\rotatebox{-270}{\makebox(0,0){\strut{}$d\sigma/dp_T(b\bar{b})$ [$\mu$b/GeV]}}}%
      \put(2680,154){\makebox(0,0){\strut{}$p_T(b\bar{b})$ [GeV]}}%
      \put(2681,1841){\makebox(0,0)[l]{\strut{}$g^*g^*\to b\bar{b}g$}}%
    }%
    \gplgaddtomacro\gplfronttext{%
      \csname LTb\endcsname%
      \put(3296,3057){\makebox(0,0)[r]{\strut{}$Q_0=4$ GeV}}%
      \csname LTb\endcsname%
      \put(3296,2771){\makebox(0,0)[r]{\strut{}$Q_0=0$ GeV}}%
    }%
    \gplbacktext
    \put(0,0){\includegraphics{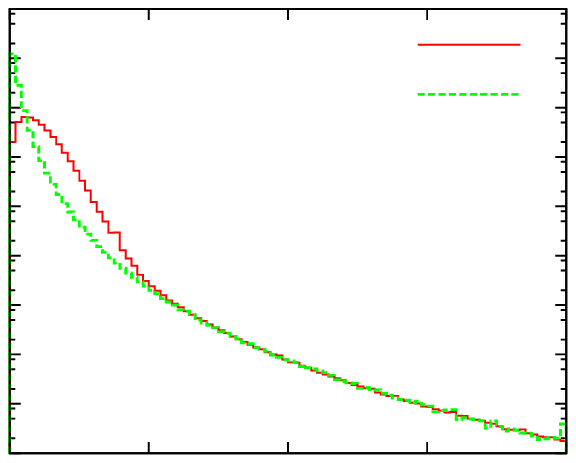}}%
    \gplfronttext
  \end{picture}%
\endgroup

\end{minipage}
\begin{minipage}[r]{0.49\linewidth}
\begingroup
  \makeatletter
  \providecommand\color[2][]{%
    \GenericError{(gnuplot) \space\space\space\@spaces}{%
      Package color not loaded in conjunction with
      terminal option `colourtext'%
    }{See the gnuplot documentation for explanation.%
    }{Either use 'blacktext' in gnuplot or load the package
      color.sty in LaTeX.}%
    \renewcommand\color[2][]{}%
  }%
  \providecommand\includegraphics[2][]{%
    \GenericError{(gnuplot) \space\space\space\@spaces}{%
      Package graphicx or graphics not loaded%
    }{See the gnuplot documentation for explanation.%
    }{The gnuplot epslatex terminal needs graphicx.sty or graphics.sty.}%
    \renewcommand\includegraphics[2][]{}%
  }%
  \providecommand\rotatebox[2]{#2}%
  \@ifundefined{ifGPcolor}{%
    \newif\ifGPcolor
    \GPcolortrue
  }{}%
  \@ifundefined{ifGPblacktext}{%
    \newif\ifGPblacktext
    \GPblacktexttrue
  }{}%
  \let\gplgaddtomacro\g@addto@macro
  \gdef\gplbacktext{}%
  \gdef\gplfronttext{}%
  \makeatother
  \ifGPblacktext
    \def\colorrgb#1{}%
    \def\colorgray#1{}%
  \else
    \ifGPcolor
      \def\colorrgb#1{\color[rgb]{#1}}%
      \def\colorgray#1{\color[gray]{#1}}%
      \expandafter\def\csname LTw\endcsname{\color{white}}%
      \expandafter\def\csname LTb\endcsname{\color{black}}%
      \expandafter\def\csname LTa\endcsname{\color{black}}%
      \expandafter\def\csname LT0\endcsname{\color[rgb]{1,0,0}}%
      \expandafter\def\csname LT1\endcsname{\color[rgb]{0,1,0}}%
      \expandafter\def\csname LT2\endcsname{\color[rgb]{0,0,1}}%
      \expandafter\def\csname LT3\endcsname{\color[rgb]{1,0,1}}%
      \expandafter\def\csname LT4\endcsname{\color[rgb]{0,1,1}}%
      \expandafter\def\csname LT5\endcsname{\color[rgb]{1,1,0}}%
      \expandafter\def\csname LT6\endcsname{\color[rgb]{0,0,0}}%
      \expandafter\def\csname LT7\endcsname{\color[rgb]{1,0.3,0}}%
      \expandafter\def\csname LT8\endcsname{\color[rgb]{0.5,0.5,0.5}}%
    \else
      \def\colorrgb#1{\color{black}}%
      \def\colorgray#1{\color[gray]{#1}}%
      \expandafter\def\csname LTw\endcsname{\color{white}}%
      \expandafter\def\csname LTb\endcsname{\color{black}}%
      \expandafter\def\csname LTa\endcsname{\color{black}}%
      \expandafter\def\csname LT0\endcsname{\color{black}}%
      \expandafter\def\csname LT1\endcsname{\color{black}}%
      \expandafter\def\csname LT2\endcsname{\color{black}}%
      \expandafter\def\csname LT3\endcsname{\color{black}}%
      \expandafter\def\csname LT4\endcsname{\color{black}}%
      \expandafter\def\csname LT5\endcsname{\color{black}}%
      \expandafter\def\csname LT6\endcsname{\color{black}}%
      \expandafter\def\csname LT7\endcsname{\color{black}}%
      \expandafter\def\csname LT8\endcsname{\color{black}}%
    \fi
  \fi
  \setlength{\unitlength}{0.0500bp}%
  \begin{picture}(4680.00,3528.00)%
    \gplgaddtomacro\gplbacktext{%
      \csname LTb\endcsname%
      \put(946,704){\makebox(0,0)[r]{\strut{} 0}}%
      \put(946,1131){\makebox(0,0)[r]{\strut{} 20}}%
      \put(946,1557){\makebox(0,0)[r]{\strut{} 40}}%
      \put(946,1984){\makebox(0,0)[r]{\strut{} 60}}%
      \put(946,2410){\makebox(0,0)[r]{\strut{} 80}}%
      \put(946,2837){\makebox(0,0)[r]{\strut{} 100}}%
      \put(946,3263){\makebox(0,0)[r]{\strut{} 120}}%
      \put(1078,484){\makebox(0,0){\strut{}-6}}%
      \put(1612,484){\makebox(0,0){\strut{}-4}}%
      \put(2146,484){\makebox(0,0){\strut{}-2}}%
      \put(2681,484){\makebox(0,0){\strut{} 0}}%
      \put(3215,484){\makebox(0,0){\strut{} 2}}%
      \put(3749,484){\makebox(0,0){\strut{} 4}}%
      \put(4283,484){\makebox(0,0){\strut{} 6}}%
      \put(176,1983){\rotatebox{-270}{\makebox(0,0){\strut{}$d\sigma/dy(b\bar{b})$ [$\mu$b]}}}%
      \put(2680,154){\makebox(0,0){\strut{}$y(b\bar{b})$}}%
    }%
    \gplgaddtomacro\gplfronttext{%
    }%
    \gplbacktext
    \put(0,0){\includegraphics{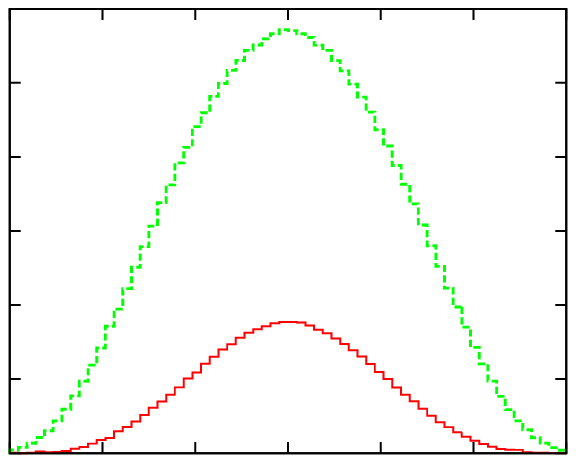}}%
    \gplfronttext
  \end{picture}%
\endgroup

\end{minipage}
\caption{\label{fig:bBg}Differential cross sections of the transverse momentum $p_T$ and the rapidity $y$ of the $b\bar{b}$-pair in the process $g^*g^*\to b\bar{b}g$. The different curves correspond to the different values of $Q_0$ in \Equation{eqn:defUnint}, with $Q_0=0$ refering to collinear factorization. The total cross sections are
$695.41\pm0.32\,\mu\mathrm{b}$ for $Q_0=0\,\GeV$,
$171.74\pm0.20\,\mu\mathrm{b}$ for $Q_0=4\,\GeV$.}
}

\myFigure{%
\begin{minipage}[l]{0.49\linewidth}
\begingroup
  \makeatletter
  \providecommand\color[2][]{%
    \GenericError{(gnuplot) \space\space\space\@spaces}{%
      Package color not loaded in conjunction with
      terminal option `colourtext'%
    }{See the gnuplot documentation for explanation.%
    }{Either use 'blacktext' in gnuplot or load the package
      color.sty in LaTeX.}%
    \renewcommand\color[2][]{}%
  }%
  \providecommand\includegraphics[2][]{%
    \GenericError{(gnuplot) \space\space\space\@spaces}{%
      Package graphicx or graphics not loaded%
    }{See the gnuplot documentation for explanation.%
    }{The gnuplot epslatex terminal needs graphicx.sty or graphics.sty.}%
    \renewcommand\includegraphics[2][]{}%
  }%
  \providecommand\rotatebox[2]{#2}%
  \@ifundefined{ifGPcolor}{%
    \newif\ifGPcolor
    \GPcolortrue
  }{}%
  \@ifundefined{ifGPblacktext}{%
    \newif\ifGPblacktext
    \GPblacktexttrue
  }{}%
  \let\gplgaddtomacro\g@addto@macro
  \gdef\gplbacktext{}%
  \gdef\gplfronttext{}%
  \makeatother
  \ifGPblacktext
    \def\colorrgb#1{}%
    \def\colorgray#1{}%
  \else
    \ifGPcolor
      \def\colorrgb#1{\color[rgb]{#1}}%
      \def\colorgray#1{\color[gray]{#1}}%
      \expandafter\def\csname LTw\endcsname{\color{white}}%
      \expandafter\def\csname LTb\endcsname{\color{black}}%
      \expandafter\def\csname LTa\endcsname{\color{black}}%
      \expandafter\def\csname LT0\endcsname{\color[rgb]{1,0,0}}%
      \expandafter\def\csname LT1\endcsname{\color[rgb]{0,1,0}}%
      \expandafter\def\csname LT2\endcsname{\color[rgb]{0,0,1}}%
      \expandafter\def\csname LT3\endcsname{\color[rgb]{1,0,1}}%
      \expandafter\def\csname LT4\endcsname{\color[rgb]{0,1,1}}%
      \expandafter\def\csname LT5\endcsname{\color[rgb]{1,1,0}}%
      \expandafter\def\csname LT6\endcsname{\color[rgb]{0,0,0}}%
      \expandafter\def\csname LT7\endcsname{\color[rgb]{1,0.3,0}}%
      \expandafter\def\csname LT8\endcsname{\color[rgb]{0.5,0.5,0.5}}%
    \else
      \def\colorrgb#1{\color{black}}%
      \def\colorgray#1{\color[gray]{#1}}%
      \expandafter\def\csname LTw\endcsname{\color{white}}%
      \expandafter\def\csname LTb\endcsname{\color{black}}%
      \expandafter\def\csname LTa\endcsname{\color{black}}%
      \expandafter\def\csname LT0\endcsname{\color{black}}%
      \expandafter\def\csname LT1\endcsname{\color{black}}%
      \expandafter\def\csname LT2\endcsname{\color{black}}%
      \expandafter\def\csname LT3\endcsname{\color{black}}%
      \expandafter\def\csname LT4\endcsname{\color{black}}%
      \expandafter\def\csname LT5\endcsname{\color{black}}%
      \expandafter\def\csname LT6\endcsname{\color{black}}%
      \expandafter\def\csname LT7\endcsname{\color{black}}%
      \expandafter\def\csname LT8\endcsname{\color{black}}%
    \fi
  \fi
  \setlength{\unitlength}{0.0500bp}%
  \begin{picture}(4680.00,3528.00)%
    \gplgaddtomacro\gplbacktext{%
      \csname LTb\endcsname%
      \put(946,704){\makebox(0,0)[r]{\strut{}$10^{-5}$}}%
      \put(946,1070){\makebox(0,0)[r]{\strut{}$10^{-4}$}}%
      \put(946,1435){\makebox(0,0)[r]{\strut{}$10^{-3}$}}%
      \put(946,1801){\makebox(0,0)[r]{\strut{}$10^{-2}$}}%
      \put(946,2166){\makebox(0,0)[r]{\strut{}$10^{-1}$}}%
      \put(946,2532){\makebox(0,0)[r]{\strut{}$10^{0}$}}%
      \put(946,2897){\makebox(0,0)[r]{\strut{}$10^{1}$}}%
      \put(946,3263){\makebox(0,0)[r]{\strut{}$10^{2}$}}%
      \put(1078,484){\makebox(0,0){\strut{} 0}}%
      \put(1879,484){\makebox(0,0){\strut{} 100}}%
      \put(2681,484){\makebox(0,0){\strut{} 200}}%
      \put(3482,484){\makebox(0,0){\strut{} 300}}%
      \put(4283,484){\makebox(0,0){\strut{} 400}}%
      \put(176,1983){\rotatebox{-270}{\makebox(0,0){\strut{}$d\sigma/dp_T(b\bar{b})$ [$\mu$b/GeV]}}}%
      \put(2680,154){\makebox(0,0){\strut{}$p_T(b\bar{b})$ [GeV]}}%
      \put(2681,2056){\makebox(0,0)[l]{\strut{}$g^*g^*\to b\bar{b}gg$}}%
    }%
    \gplgaddtomacro\gplfronttext{%
      \csname LTb\endcsname%
      \put(3296,3057){\makebox(0,0)[r]{\strut{}$Q_0=4$ GeV}}%
      \csname LTb\endcsname%
      \put(3296,2771){\makebox(0,0)[r]{\strut{}$Q_0=0$ GeV}}%
    }%
    \gplbacktext
    \put(0,0){\includegraphics{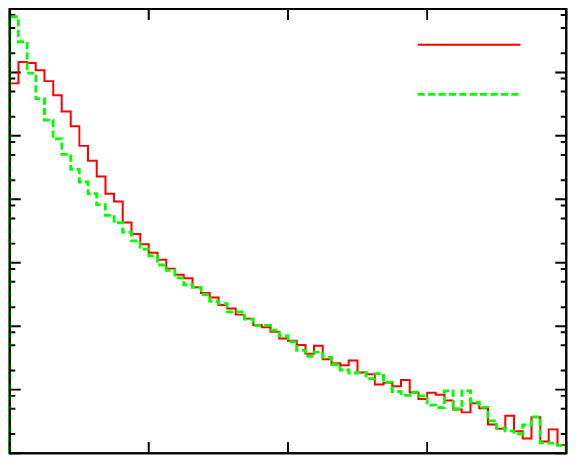}}%
    \gplfronttext
  \end{picture}%
\endgroup

\end{minipage}
\begin{minipage}[r]{0.49\linewidth}
\begingroup
  \makeatletter
  \providecommand\color[2][]{%
    \GenericError{(gnuplot) \space\space\space\@spaces}{%
      Package color not loaded in conjunction with
      terminal option `colourtext'%
    }{See the gnuplot documentation for explanation.%
    }{Either use 'blacktext' in gnuplot or load the package
      color.sty in LaTeX.}%
    \renewcommand\color[2][]{}%
  }%
  \providecommand\includegraphics[2][]{%
    \GenericError{(gnuplot) \space\space\space\@spaces}{%
      Package graphicx or graphics not loaded%
    }{See the gnuplot documentation for explanation.%
    }{The gnuplot epslatex terminal needs graphicx.sty or graphics.sty.}%
    \renewcommand\includegraphics[2][]{}%
  }%
  \providecommand\rotatebox[2]{#2}%
  \@ifundefined{ifGPcolor}{%
    \newif\ifGPcolor
    \GPcolortrue
  }{}%
  \@ifundefined{ifGPblacktext}{%
    \newif\ifGPblacktext
    \GPblacktexttrue
  }{}%
  \let\gplgaddtomacro\g@addto@macro
  \gdef\gplbacktext{}%
  \gdef\gplfronttext{}%
  \makeatother
  \ifGPblacktext
    \def\colorrgb#1{}%
    \def\colorgray#1{}%
  \else
    \ifGPcolor
      \def\colorrgb#1{\color[rgb]{#1}}%
      \def\colorgray#1{\color[gray]{#1}}%
      \expandafter\def\csname LTw\endcsname{\color{white}}%
      \expandafter\def\csname LTb\endcsname{\color{black}}%
      \expandafter\def\csname LTa\endcsname{\color{black}}%
      \expandafter\def\csname LT0\endcsname{\color[rgb]{1,0,0}}%
      \expandafter\def\csname LT1\endcsname{\color[rgb]{0,1,0}}%
      \expandafter\def\csname LT2\endcsname{\color[rgb]{0,0,1}}%
      \expandafter\def\csname LT3\endcsname{\color[rgb]{1,0,1}}%
      \expandafter\def\csname LT4\endcsname{\color[rgb]{0,1,1}}%
      \expandafter\def\csname LT5\endcsname{\color[rgb]{1,1,0}}%
      \expandafter\def\csname LT6\endcsname{\color[rgb]{0,0,0}}%
      \expandafter\def\csname LT7\endcsname{\color[rgb]{1,0.3,0}}%
      \expandafter\def\csname LT8\endcsname{\color[rgb]{0.5,0.5,0.5}}%
    \else
      \def\colorrgb#1{\color{black}}%
      \def\colorgray#1{\color[gray]{#1}}%
      \expandafter\def\csname LTw\endcsname{\color{white}}%
      \expandafter\def\csname LTb\endcsname{\color{black}}%
      \expandafter\def\csname LTa\endcsname{\color{black}}%
      \expandafter\def\csname LT0\endcsname{\color{black}}%
      \expandafter\def\csname LT1\endcsname{\color{black}}%
      \expandafter\def\csname LT2\endcsname{\color{black}}%
      \expandafter\def\csname LT3\endcsname{\color{black}}%
      \expandafter\def\csname LT4\endcsname{\color{black}}%
      \expandafter\def\csname LT5\endcsname{\color{black}}%
      \expandafter\def\csname LT6\endcsname{\color{black}}%
      \expandafter\def\csname LT7\endcsname{\color{black}}%
      \expandafter\def\csname LT8\endcsname{\color{black}}%
    \fi
  \fi
  \setlength{\unitlength}{0.0500bp}%
  \begin{picture}(4680.00,3528.00)%
    \gplgaddtomacro\gplbacktext{%
      \csname LTb\endcsname%
      \put(946,704){\makebox(0,0)[r]{\strut{} 0}}%
      \put(946,1070){\makebox(0,0)[r]{\strut{} 20}}%
      \put(946,1435){\makebox(0,0)[r]{\strut{} 40}}%
      \put(946,1801){\makebox(0,0)[r]{\strut{} 60}}%
      \put(946,2166){\makebox(0,0)[r]{\strut{} 80}}%
      \put(946,2532){\makebox(0,0)[r]{\strut{} 100}}%
      \put(946,2897){\makebox(0,0)[r]{\strut{} 120}}%
      \put(946,3263){\makebox(0,0)[r]{\strut{} 140}}%
      \put(1078,484){\makebox(0,0){\strut{}-6}}%
      \put(1612,484){\makebox(0,0){\strut{}-4}}%
      \put(2146,484){\makebox(0,0){\strut{}-2}}%
      \put(2681,484){\makebox(0,0){\strut{} 0}}%
      \put(3215,484){\makebox(0,0){\strut{} 2}}%
      \put(3749,484){\makebox(0,0){\strut{} 4}}%
      \put(4283,484){\makebox(0,0){\strut{} 6}}%
      \put(176,1983){\rotatebox{-270}{\makebox(0,0){\strut{}$d\sigma/dy(b\bar{b})$ [$\mu$b]}}}%
      \put(2680,154){\makebox(0,0){\strut{}$y(b\bar{b})$}}%
    }%
    \gplgaddtomacro\gplfronttext{%
    }%
    \gplbacktext
    \put(0,0){\includegraphics{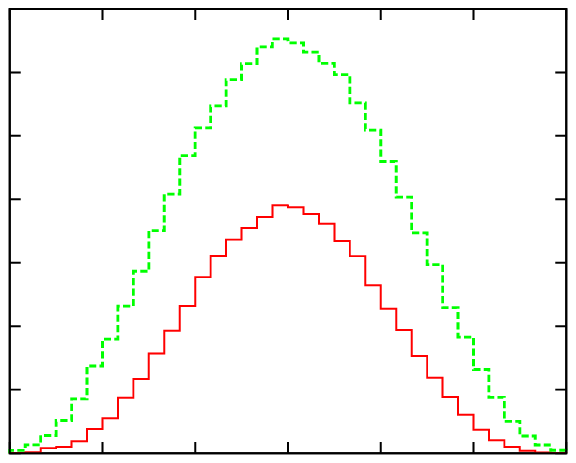}}%
    \gplfronttext
  \end{picture}%
\endgroup

\end{minipage}
\caption{\label{fig:bBgg}Differential cross sections of the transverse momentum $p_T$ and the rapidity $y$ of the $b\bar{b}$-pair in the process $g^*g^*\to b\bar{b}gg$. The different curves correspond to the different values of $Q_0$ in \Equation{eqn:defUnint}, with $Q_0=0$ refering to collinear factorization. The total cross sections are
$772.4\pm1.2\,\mu\mathrm{b}$ for $Q_0=0\,\GeV$,
$393.9\pm1.2\,\mu\mathrm{b}$ for $Q_0=4\,\GeV$.}
}


In \Figure{fig:bBlL}, \Figure{fig:bBZg}, \Figure{fig:bBg} and \Figure{fig:bBgg} some distributions are depicted for the mentioned processes.
Results for two values of the scale $Q_0$ in \Equation{eqn:defUnint} are presented: $Q_0=0$, which refers to the result within collinear factorization, and $Q_0=4\,\GeV$.
The latter value was chosen such that it leads to a ratio of the cross sections for the process  $g^*g^*\to b\bar{b}Z$ similar to those presented in \cite{Deak:2008ky}.
The absolute values are considerably smaller, because we present results for an off-shell $Z$-boson with only the $\mu^+\mu^-$-channel taken into account.

We want to stress that the presented results should merely be seen as a proof of the computational potential of our approach.
Still, we may state that the results are consistent with those presented in \eg~\cite{Deak:2008ky,Baranov:2008hj}.
The cross sections within high-energy factorization are suppressed compared to collinear factorization at leading-order QCD, and the transversal-momentum distributions show a larger spread.

%
\section{Summary\label{Sec:conclusion}}
We presented a prescription to calculate tree-level helicity amplitudes for arbitrary scattering processes with off-shell initial-state gluons within the kinematics of high-energy scattering.
We showed that the amplitudes are manifestly gauge invariant, and that the approach is equivalent to Lipatov's effective action approach.
We established the connection with earlier work regarding multi-gluon amplitudes with one off-shell initial-state gluon, and we showed the computational potential of the presented approach with numerical calculations for scattering processes with several particles in the final state.
%

\subsection*{Acknowledgments}
This work was partially supported by HOMING PLUS/2010-2/6: ``Matrix Elements and Exclusive Parton Densities for Large Hadron Collider''.

\newpage
\bibliography{bibliography}{}
\bibliographystyle{JHEP}


\begin{appendix}
\section{\label{App:spinors}Definition of spinors}
\newcommand{\Rsp}{R}
\newcommand{\RspS}{R_{*}}
\newcommand{\Lsp}{L}
\newcommand{\LspS}{L_{*}}
\newcommand{\Trans}{T}
\newcommand{\zerotwo}{\mathbf{0}}
\newcommand{\unittwo}{\mathbf{1}}
\newcommand{\unitfour}{\mathbf{1}_{4\times4}}
\newcommand{\vectwo}[2]{\begin{pmatrix}#1\\#2\end{pmatrix}}
\newcommand{\mattwo}[4]{\begin{pmatrix}#1&#2\\#3&#4\end{pmatrix}}
We introduce the bi-spinor of a four-momentum $p$ following
%
\begin{align}
\lid{\sigma}{p} = \sigma_0p_0 - \lid{\vec{\sigma}}{\vec{p}} = \mattwo{p_0-p_3}{-p_1+\imag p_2}{-p_1-\imag p_2}{p_0+p_3}
~,
\end{align}
%
where $\sigma_0$ is the identity $2\times2$-matrix and $\vec{\sigma}=(\sigma_1,\sigma_2,\sigma_3)$ are the Pauli matrices.
We also define
%
\begin{equation}
\lid{\tilde{\sigma}}{p} = \sigma_0p_0 + \lid{\vec{\sigma}}{\vec{p}}
~.
\end{equation}
For $p^2=p_0^2-p_1^2-p_2^2-p_3^2=0$, we may write
%
\begin{align}
\lid{\sigma}{p} = \Lsp(p)\,\LspS(p)^{\Trans}
\quad,\quad
\lid{\tilde{\sigma}}{p} = \Rsp(p)\,\RspS(p)^{\Trans}
~,
\end{align}
with
\begin{align}
\Lsp(p) &= \frac{1}{\sqrt{|p_0+p_3|}}
         \vectwo{-p_1+\imag p_2}{p_0+p_3 }
&
\LspS(p) &= \frac{\sqrt{|p_0+p_3|}}{p_0+p_3}
         \vectwo{-p_1-\imag p_2}{p_0+p_3 }
\nonumber\\
\Rsp(p) &= \frac{\sqrt{|p_0+p_3|}}{p_0+p_3}
         \vectwo{p_0+p_3}{p_1+\imag p_2}
&
\RspS(p) &= \frac{1}{\sqrt{|p_0+p_3|}}
         \vectwo{p_0+p_3}{p_1-\imag p_2}
~.
\end{align}
%
If $p$ is real with $p_0>0$, then $\LspS(p)=\Lsp(p)^{*}$ and $\RspS(p)=\Rsp(p)^{*}$, and $\Lsp(p),\Rsp(p)$ are the well-known Weyl spinors.
The spinors above are defined also for complex momentum, as long as it is on-shell.
Introducing the bra-ket-notation, we write
\begin{align}
\ketm{p}&=\vectwo{\Lsp(p)}{\zerotwo}
\quad,\quad
\bram{p}=\big(\,\zerotwo\,,\,\LspS(p)^\Trans\,\big)
\notag\\
\ketp{p}&=\vectwo{\zerotwo}{\Rsp(p)}
\quad,\quad
\brap{p}=\big(\,\RspS(p)^\Trans\,,\,\zerotwo\,\big)
\end{align}
so that
\begin{equation}
\ketp{p}\brap{p} \,+\, \ketm{p}\bram{p}
=\mattwo{\zerotwo}{\lid{\sigma}{p}}{\lid{\tilde{\sigma}}{p}}{\zerotwo}
=\slashp
\label{Eq04}
~.
\end{equation}
This leads to the Weyl representation of the $\gamma$-algebra, with
%
\begin{equation}
\gamma_5
=\imag\gamma^0\gamma^1\gamma^2\gamma^3
= \mattwo{-\unittwo}{\zerotwo}{\zerotwo}{\unittwo}
~.
\end{equation}
%
We state some basic relations.
We trivially have
%
\begin{equation}
\bkpp{p}{p}=\bkmm{p}{p}=0
~,
\end{equation}
%
and explicit calculation reveals that also
%
\begin{equation}
\bkmp{p}{p}=\bkpm{p}{p}=0
~.
\end{equation}
%
From explicit calculation also follows that
%
\begin{equation}
\bkmp{p}{q}\,\bkpm{q}{p}=2\,\lid{p}{q}
~.
\end{equation}
%
One more important relation is that
\begin{align}
2\,\lid{p}{q}
&= \bkmp{p}{q}\bkpm{q}{p} \,+\, \bkmm{p}{q}\bkmm{q}{p}
\notag\\
&= \bram{p}\,\big(\,\ketp{q}\brap{q} \,+\, \ketm{q}\bram{q}\,\big)\,\ketm{p}
\notag\\
&= \bram{p}\,\slashq\,\ketm{p}
~.
\end{align}
By decomposing any arbitrary four-vector into two light-like vector, it can easily be shown that this relation also holds if $q$ is not light-like.

Realize, finally, that in explicit numerical calculations, it is usually not wise to identify $p_1=p_x, p_2=p_y, p_3=p_z$, since in collider physics one may typically encounter situations in which $p_0=-p_z$.
Therefore, it is better to identify \eg\ $p_1=p_z, p_2=p_x, p_3=p_y$.

\section{\label{App:Lipatov}Explict calculation of reggeon-gluon vertices}
In this appendix, we calculate some simple reggeon-gluon vertices using our prescription.
The calculations are performed in the Feynman gauge.
\subsection{The reggeon-gluon-gluon vertex}
First, consider the process $g^*g\to g$.
The process is embedded in 
%
\begin{equation}
q_A(p_A,j)\,g(p_2,a,\nu)\to q_A(p_{A'},i)\,g(k_1+p_2,b,\sigma)
~.\nonumber
\end{equation}
%
The arguments $a$ and $b$ refer to the color of the gluons, the arguments $\nu$ and $\sigma$ are the Lorentz indices of the gluons, and the arguments $j$ and $i$ refer to the color of the quarks.
The necessary Feynman graphs are depicted in \Figure{Fig3}.
According to the prescription, the fermion line should be interpreted as an eikonal line with momenta $p_A=k_1$ and $p_{A'}=0$.
The sum $C$ of all three graphs is
%
\begin{multline}
C=
\imag\gqcd\xone\sqrt{-2k_1^2}\bigg[
T_{ij}^c\,\ell_1^\mu\,\frac{-\imag}{k_1^2}\,f^{cab}\,{V_{\mu}}^{\nu\sigma}(k_1,p_2,-k_1-p_2)
\\
 + \frac{(T^aT^b)_{ij}\,\ell_1^\nu\ell_1^\sigma}{\lid{\ell_1}{(k_1+p_2)}}
 + \frac{(T^bT^a)_{ij}\,\ell_1^\sigma\ell_1^\nu}{\lid{\ell_1}{(-p_2)}}
\bigg]
~,\nonumber
\end{multline}
where $V^{\mu\nu\sigma}$ is the usual vertex for three vector-bosons
%
\begin{equation}
V^{\mu\nu\sigma}(p_1,p_2,p_3)
=(p_1-p_2)^\sigma\,g^{\mu\nu} + (p_2-p_3)^\mu\,g^{\nu\sigma} + (p_3-p_1)^\nu\,g^{\sigma\mu}
~.
\end{equation}
%
Using $\lid{\ell_1}{k_1}=0$, we find
\begin{multline}
C=
\imag\gqcd\xone\sqrt{-2k_1^2}\bigg[
\frac{-\imag\,T_{ij}^c\,f^{cab}}{k_1^2}\big[
2(\lid{\ell_1}{p_2})g^{\nu\sigma} -(2k_1+p_2)^\nu\ell_1^\sigma +(k_1-p_2)^\sigma\ell_1^\nu
\big]
\\
 + \frac{[T^a,T^b]_{ij}}{\lid{\ell_1}{p_2}}\,\ell_1^\nu\ell_1^\sigma
\bigg]
~.\nonumber
\end{multline}
Inserting $[T^a,T^b]=\imag f^{cab}T^c$, we get
\begin{align}
C&=
\frac{-\gqcd\xone\,\sqrt{2}T_{ij}^c\,f^{cab}}{\sqrt{-k_1^2}}\left[
2(\lid{\ell_1}{p_2})g^{\nu\sigma} -(2k_1+p_2)^\nu\ell_1^\sigma +(k_1-p_2)^\sigma\ell_1^\nu
-\frac{k_1^2}{\lid{\ell_1}{p_2}}\,\ell_1^\nu\ell_1^\sigma
\right]
\nonumber
\notag\\
&=
\frac{-\xone E}{\sqrt{-k_1^2}}\,\sqrt{2}T^c_{ij}
\;\gamma_{\parallel}^{\nu\sigma-}(p_2,a;k_1+p_2,b;k_1,c)
~,\notag
\end{align}
%
with
%
\begin{multline}
\gamma_{\parallel}^{\nu\sigma-}(p_2,a;p_1,b;k_1,c)\\
=
\frac{\gqcd}{E}\,f^{abc}\left[
2(\lid{\ell_1}{p_2})g^{\nu\sigma} -(2p_1-p_2)^\nu\ell_1^\sigma -(2p_2-p_1)^\sigma\ell_1^\nu
-\frac{k_1^2}{\lid{\ell_1}{p_2}}\,\ell_1^\nu\ell_1^\sigma
\right]
\end{multline}
%
which, for $\ell_1=E\,n^{-}$, is identical to equation~(34) in \cite{Antonov:2004hh}.

\subsection{The reggeon-reggeon-gluon vertex}
\myFigure{%
\epsfig{figure=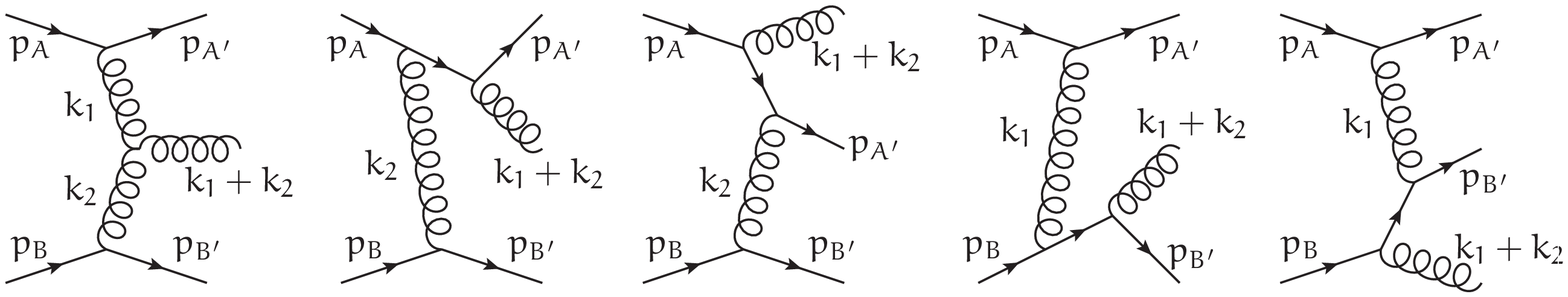,width=0.95\linewidth}
\caption{\label{Fig9}Feynman graphs for $q_A\,q_B\to q_A\,q_B\,g$.}
}
Next, we consider the process $g^*g^*\to g$.
It is embedded in
%
\begin{equation}
q_A(p_A,j)\,q_B(p_B,l) \to q_A(p_{A'},i)\,q_B(p_{B'},k)\,g(k_1+k_2,b,\sigma)
~,\nonumber
\end{equation}
%
where the arguments $j$, $l$, $i$ and $k$ refer to the color of the quarks, the argument $b$ refers to the color of the gluon, and $\sigma$ is the Lorentz index of the gluon.
The necessary graphs are depicted in \Figure{Fig9}.
According to our prescription, we have $p_A=k_1$, $p_B=k_2$, $p_{A'}=p_{B'}=0$, and the sum of the graphs is equal to
%
\begin{multline}
C=
-2\gqcd\xone\xtwo\sqrt{k_1^2k_2^2}\bigg[
T_{ij}^c\,\ell_1^\mu\,\frac{-\imag}{k_1^2}\,f^{cab}\,{{V_{\mu\nu}}^\sigma}(k_1,k_2,-k_1-k_2)\,\frac{-\imag}{k_2^2}\,T^a_{kl}\,\ell_2^\nu
\\
\hspace{80pt}
+\frac{(T^aT^b)_{ij}\,\ell_1^\nu\ell_1^\sigma}{\lid{\ell_1}{(k_1+k_2)}}\,\frac{-\imag}{k_2^2}\,T^a_{kl}\,\ell_{2\nu}
+\frac{(T^bT^a)_{ij}\,\ell_1^\sigma\ell_1^\nu}{\lid{\ell_1}{(-k_2)}}\,\frac{-\imag}{k_2^2}\,T^a_{kl}\,\ell_{2\nu}
\\
+T^c_{ij}\,\ell_{1\mu}\,\frac{-\imag}{k_1^2}
\,\frac{(T^cT^b)_{kl}\,\ell_2^\mu\ell_2^\sigma}{\lid{\ell_2}{(k_1+k_2)}}
+T^c_{ij}\,\ell_{1\mu}\,\frac{-\imag}{k_1^2}
\,\frac{(T^bT^c)_{kl}\,\ell_2^\sigma\ell_2^\mu}{\lid{\ell_2}{(-k_1)}}
\bigg]
\nonumber
\end{multline}
Using $\lid{\ell_1}{k_1}=\lid{\ell_2}{k_2}=0$, we find
\begin{multline}
C=
-2\gqcd\xone\xtwo\sqrt{k_1^2k_2^2}\bigg[
\frac{-T_{ij}^cT^a_{kl}f^{cab}}{k_1^2k_2^2}\big[
2(\lid{\ell_1}{k_2})\ell_2^\sigma -2(\lid{\ell_2}{k_1})\ell_1^\sigma +(\lid{\ell_1}{\ell_2})(k_1-k_2)^\sigma
\\
+\frac{[T^a,T^b]_{ij}\,\ell_1^\nu\ell_1^\sigma}{\lid{\ell_1}{k_2}}\,\frac{-\imag}{k_2^2}\,T^a_{kl}\,\ell_{2\nu}
+T^c_{ij}\,\ell_{1\mu}\,\frac{-\imag}{k_1^2}
\,\frac{[T^c,T^b]_{kl}\,\ell_2^\mu\ell_2^\sigma}{\lid{\ell_2}{k_1}}
\bigg]
~.\nonumber
\end{multline}
Inserting $[T^a,T^b]=\imag f^{abc}T^c$ and $[T^c,T^b]=\imag f^{cba}T^a$, we get
\begin{multline}
C
=-2\gqcd\xone\xtwo\sqrt{k_1^2k_2^2}\bigg[
\frac{-T_{ij}^cT^a_{kl}f^{cab}}{k_1^2k_2^2}\big[
2(\lid{\ell_1}{k_2})\ell_2^\sigma -2(\lid{\ell_2}{k_1})\ell_1^\sigma +(\lid{\ell_1}{\ell_2})(k_1-k_2)^\sigma
\big]
\\
+\frac{f^{abc}T^c_{ij}T^a_{kl}}{(\lid{\ell_1}{k_2})k_2^2}\,(\lid{\ell_1}{\ell_2})\ell_1^\sigma
+\frac{f^{cba}T^a_{kl}T^c_{ij}}{(\lid{\ell_2}{k_1})k_1^2}\,(\lid{\ell_1}{\ell_2})\ell_2^\sigma
\bigg]
~.\nonumber
\end{multline}
Using the anti-symmetry of $f^{abc}$, we finally find
\begin{align}
C
&
=\frac{2\gqcd\xone\xtwo\lid{\ell_1}{\ell_2}}{\sqrt{k_1^2k_2^2}}\,T^c_{ij}T^a_{kl}f^{cba}\bigg[
-2\frac{\lid{\ell_1}{k_2}}{\lid{\ell_1}{\ell_2}}\ell_2^\sigma
+2\frac{\lid{\ell_2}{k_1}}{\lid{\ell_1}{\ell_2}}\ell_1^\sigma
-(k_1-k_2)^\sigma
+\frac{k_1^2}{\lid{\ell_1}{k_2}}\ell_1^\sigma
-\frac{k_2^2}{\lid{\ell_2}{k_1}}\ell_2^\sigma
\bigg]
\notag\\
&
=\frac{\xone\xtwo E^2}{\sqrt{k_1^2k_2^2}}\,2T^c_{ij}T^a_{kl}
\,\Gamma^{-\sigma+}(k_1,c;k_1+k_2,b;k_2,a)
~,\notag
\end{align}
%
with
%
\begin{multline}
\Gamma^{-\sigma+}(k_1,c;k_1+k_2,b;k_2,a)\\
=
2\,\gqcd\,f^{abc}\left[
\left(\frac{\lid{\ell_2}{k_1}}{E^2}+\frac{k_1^2}{\lid{\ell_1}{k_2}}\right)\ell_1^\sigma
-\left(\frac{\lid{\ell_1}{k_2}}{E^2}+\frac{k_2^2}{\lid{\ell_2}{k_1}}\right)\ell_2^\sigma
+(k_2-k_1)^\sigma
\right]
\end{multline}
%
which, for $\ell_1=E\,n^{-}$ and $\ell_2=E\,n^+$, is identical to the first line of equation~(41) in~\cite{Antonov:2004hh} combined with equation~(42) in~\cite{Antonov:2004hh}.
Realize that $\lid{\ell_1}{\ell_2}=2E^2$.

\section{\label{App:proof}Proof of \Equation{Eqn:traceident}}
Here, we proof that for a cycle-symmetric tensor $\tau^{i_{1}\ldots i_{n}}$ and the set
of the complex numbers $\xi_{1},\ldots,\xi_{n}$ satisfying $\xi_{1}+\ldots+\xi_{n}=0$
the following identity holds
\begin{equation}
\sum_{\left\{ i_{1},\ldots,i_{n}\right\} }\frac{\tau^{i_{1}\ldots i_{n}}}{\xi_{i_{1}}\left(\xi_{i_{1}}+\xi_{i_{2}}\right)\ldots\left(\xi_{i_{1}}+\xi_{i_{2}}+\ldots+\xi_{i_{n-1}}\right)}=0~,
\end{equation}
where the sum goes over all the permutations of the indices $i_{1},\ldots,i_{n}$.

Let us decompose the sum over all permutations as follows\begin{equation}
\sum_{\left\{ i_{1},\ldots,i_{n}\right\} }\frac{\tau^{i_{1}\ldots i_{n}}}{\xi_{i_{1}}\left(\xi_{i_{1}}+\xi_{i_{2}}\right)\ldots\left(\xi_{i_{1}}+\xi_{i_{2}}+\ldots+\xi_{i_{n-1}}\right)}=\sum_{\left\{ i_{1},\ldots,i_{n}\right\} '}\tau^{i_{1}\ldots i_{n}}\sum_{\left(i_{1},\ldots,i_{n}\right)}a\left(i_{1},\ldots,i_{n}\right),\label{eq:A}\end{equation}
where\begin{equation}
a\left(i_{1},\ldots,i_{n}\right)=\frac{1}{\xi_{i_{1}}\left(\xi_{i_{1}}+\xi_{i_{2}}\right)\ldots\left(\xi_{i_{1}}+\xi_{i_{2}}+\ldots+\xi_{i_{n-1}}\right)}\end{equation}
and $\left\{ i_{1},\ldots,i_{n}\right\} '$ denotes the set of all
non-cyclic permutations of the indices. The notation $\left(i_{1},\ldots,i_{n}\right)$
refers to the set of all cyclic permutations. Let us construct the
following meromorphic function\begin{equation}
f\left(\xi\right)=\frac{-1}{\left(-\xi\right)\left(-\xi+\xi_{i_{1}}\right)\left(-\xi+\xi_{i_{1}}+\xi_{i_{2}}\right)\ldots\left(-\xi+\xi_{i_{1}}+\xi_{i_{2}}+\ldots+\xi_{i_{n-1}}\right)}~.\end{equation}
The single proper poles are\begin{align}
z_{0} & =0,\nonumber \\
z_{1} & =\xi_{i_{1}},\nonumber \\
z_{2} & =\xi_{i_{1}}+\xi_{i_{2}},\nonumber \\
 & \vdots\nonumber \\
z_{n} & =\xi_{i_{1}}+\xi_{i_{2}}+\ldots+\xi_{i_{n-1}}.\end{align}
The residues of $f\left(\xi\right)$ in those points are\begin{equation}
\underset{\xi=z_{0}}{\mathrm{res}}\, f\left(\xi\right)=\frac{1}{\xi_{i_{1}}\left(\xi_{i_{1}}+\xi_{i_{2}}\right)\ldots\left(\xi_{i_{1}}+\xi_{i_{2}}+\ldots+\xi_{i_{n-1}}\right)}=a\left(i_{1},\ldots,i_{n}\right),\end{equation}
\begin{multline}
\underset{\xi=z_{1}}{\mathrm{res}}\, f\left(\xi\right)=\frac{1}{\left(-\xi_{i_{1}}\right)\xi_{i_{2}}\left(\xi_{i_{2}}+\xi_{i_{3}}\right)\ldots\left(\xi_{i_{2}}+\ldots+\xi_{i_{n-1}}\right)}\\
=\frac{1}{\xi_{i_{2}}\left(\xi_{i_{2}}+\xi_{i_{3}}\right)\ldots\left(\xi_{i_{2}}+\ldots+\xi_{i_{n-1}}\right)\left(\xi_{i_{2}}+\ldots+\xi_{i_{n-1}}+\xi_{i_{n}}\right)}\\
=a\left(i_{2},\ldots,i_{n},i_{1}\right),\end{multline}
\begin{multline*}
\underset{\xi=z_{2}}{\mathrm{res}}\, f\left(\xi\right)=\frac{1}{\left(-\xi_{i_{1}}-\xi_{i_{2}}\right)\left(-\xi_{i_{2}}\right)\xi_{i_{3}}\left(\xi_{i_{3}}+\xi_{i_{4}}\right)\ldots\left(\xi_{i_{2}}+\ldots+\xi_{i_{n-1}}\right)}\\
=\frac{1}{\xi_{i_{3}}\left(\xi_{i_{3}}+\xi_{i_{4}}\right)\ldots\left(\xi_{i_{3}}+\ldots+\xi_{i_{n-1}}\right)\left(\xi_{i_{3}}+\ldots+\xi_{i_{n}}\right)\left(\xi_{i_{3}}+\ldots+\xi_{i_{n}}+\xi_{i_{1}}\right)}\\
=a\left(i_{3},\ldots,i_{n},i_{1},i_{2}\right),\end{multline*}
and so on. Let us now consider the integral of $f\left(\xi\right)$
over the circle with origin $\xi=0$ and radius $R\rightarrow\infty$.
Obviously\begin{equation}
\ointop_{R\rightarrow\infty}f\left(\xi\right)d\xi=0~.\end{equation}
On the other hand we have\begin{equation}
0=\sum_{\mathrm{res}}f\left(\xi\right)=\sum_{\left(i_{1},\ldots,i_{n}\right)}a\left(i_{1},\ldots,i_{n}\right)\end{equation}
and thus the sums in Eq. (\ref{eq:A}) vanish.
Let us finally remark that we cross-checked the identity explicitly up to $n=6$ with the help of {\sc FORM}~\cite{Kuipers:2012rf}.

\end{appendix}
\end{document}